\documentclass[twocolumn,aps,pra,floatfix,longbibliography]{revtex4-2}

\usepackage{amsmath}
\usepackage{txfonts}
\usepackage{microtype}
\usepackage{graphicx}
\usepackage{color}
\usepackage{ulem}
\usepackage{bm}
\usepackage{comment}
\usepackage[utf8]{inputenc}
\usepackage[cal=pxtx]{mathalfa}

\begin{document}

 \title{Time delay in  nonadiabatic tunneling ionization}

\author{Michael Klaiber}\email{klaiber@mpi-hd.mpg.de}
\author{Karen Z. Hatsagortsyan}\email{k.hatsagortsyan@mpi-hd.mpg.de}
\author{Christoph H. Keitel}
\affiliation{Max-Planck-Institut f\"ur Kernphysik, Saupfercheckweg 1, 69117 Heidelberg, Germany}

\date{\today}

\begin{abstract}

The under-the-barrier recollision pathway during the tunneling ionization of an atom, despite its minuscule amplitude, leaves its signature on the observable photoelectron momentum distribution due to a phase shift relative to the direct ionization path, that can be interpreted as a tunneling time delay.
We investigate how this time delay is modified in the nonadiabatic regime, where the tunneling barrier evolves dynamically during the ionization process. Our analysis is based on a simple model of a one-dimensional short-range potential driven by a strong  laser pulse  which remains amenable to a systematic analytical treatment with the strong field approximation~(SFA). We show a significant contribution of the high-order  SFA corrections, which describes the recollision channel with recombination and further ionization, and is additional to the  rescattering.  Whereas the recombination channel dominates over the rescattering for the tunneling time delay in the quasistatic regime at relatively large laser fields, their contributions are reversed in weak fields.

\end{abstract}

\maketitle

\section{Introduction}

The role of sub-barrier dynamics is important for understanding strong-field ionization physics. In particular, the imaginary-time method \cite{Popov_2005}, which addresses this dynamics, was the first to derive strong-field ionization probabilities that account for Coulomb effects \cite{Perelomov_1966a, Perelomov_1967a, Perelomov_1967b, Popov_2004u}, advancing the Keldysh theory \cite{Keldysh_1965}. The phase of the electron quantum orbit accumulated during the sub-barrier motion \cite{Yan_2012a} can shift ultrafast holography patterns \cite{Huismans_2011}, while sub-barrier Coulomb effects \cite{Keil_2016, He_2022} can induce specific modifications in the photoelectron momentum distribution (PMD).

A new pathway during the tunneling ionization of an atom has been identified in Ref.~\cite{Klaiber_2018} via under-the-barrier recollisions. We demonstrated that the interference between direct and under-the-barrier recolliding quantum orbits induces subtle changes in the PMD in the quasistatic regime, which can be interpreted as a tunneling time delay. This approach provides a full quantum mechanical description of the tunneling time delay in the attoclock \cite{Eckle_2008a, Eckle_2008b, Pfeiffer_2012, Landsman_2014o, Camus_2017, Sainadh_2019}, complementing backpropagation \cite{Ni_2016, Ni_2018b, Ni_2018a}, Wigner-time, and other existing descriptions \cite{Teeny_2016a, Teeny_2016b, Yakaboylu_2013, Kheifets_2020, Eicke_2018, Douguet_2018, Bray_2018, Canario_2021, Klaiber_2022R, Klaiber_2023, Klaiber_2024}. Furthermore, we found that the well-known Freeman resonances \cite{Freeman_1987}, when an intermediate bound resonant state is involved in the multiphoton ionization, can proceed in the strong field regime via an under-the-barrier recollision path, the latter inducing a transient population of Rydberg states followed by ionization  \cite{Klaiber_2020}. Counterintuitively, this process can be significantly more efficient in strong fields than direct multiphoton transitions, as recently demonstrated experimentally \cite{Khurelbaatar_2025}.

In this paper, we investigate the time delay in strong field ionization within a large range of laser field intensities, the regime of the interaction spanning from the tunneling to multiphoton regimes. To accurately account for the time delay analytically, we employ a simple one-dimensional  model with  a short-range atomic potential, and  a half-cycle driving laser pulse.  The time-dependent Schr\"odinger equation (TDSE) is solved using the strong-field approximation (SFA) \cite{Keldysh_1965,Faisal_1973,Reiss_1980}. For  enhanced accuracy, we avoid the saddle-point approximation (SPA) in the calculation of time-integrals in the amplitudes. While similar calculations in Ref.~\cite{Klaiber_2018} for the quasistatic regime describe the under-the-barrier recollision heuristically by replacing the Born-approximation amplitude for the short-range potential scattering in the SFA matrix element with the exact $T$-matrix,  here we systematically apply high-order SFA theory, because in the nonadiabatic regime the validity of the low-frequency approximation with  the laser-free $T$-matrix is unfounded. In this framework, the under-the-barrier dynamics is described with two possible contributions, as rescattering and recollision-recombination-ionization  channels, the latter emerging naturally in the high-order SFA correction and giving a significant contribution to the tunneling time. The developed theory allows for the extension of tunneling time theory into the weak-field nonadiabatic regime (high Keldysh parameters $\gamma$) and enables a detailed analysis of its behavior. 
We emphasize the significance of the nonadiabatic regime. Serving as a bridge between multiphoton and tunneling ionization, this  intermediate regime offers broad applicability as most ionization experiments have been carried out in this regime, and hosts  intriguing strong-field ionization phenomena, see e.g. \cite{Yudin_2001b,Kaushal_2013,Boge_2013,Klaiber_2015,Dubois_2020a,Trabert_2021,Spencer_2026}. Crucially, it gives rise to key nonadiabatic effect—energy gain by the tunneling electron during the barrier passage—which generates the most intricate tunneling-time behavior, for instance, an extremely large unexplained offset angle of the attoclock at  $\gamma\approx 3.8$ \cite{Boge_2013}.

The results obtained here using a one-dimensional model remain qualitatively valid for strong-field ionization in a circularly polarized laser field, as confirmed in Ref.~ \cite{Klaiber_2018},  describing the shift of the absolute peak in the PMD within the nonadiabatic regime. Unlike in linear polarization, circular polarization produces no significant intracycle interference structures in the PMD that could obscure the momentum shift induced by the time delay.

As is known, the strong field ionization regime can be identified by two independent parameters \cite{Popruzhenko_2014r}, in particular, by the Keldysh parameter $\gamma=\omega_0 \sqrt{2I_p}/E_0$ \cite{Keldysh_1965}, and the laser frequency $\omega_0$, with the field amplitude $E_0$, and the ionization energy $I_p$. We may distinguish physically the $\omega_0 \ll I_p$  regime from the $\omega_0 \sim I_p$ regime; in the former, the excited atomic states play a minor role in the ionization dynamics, whereas in the latter, they play a major role.  Within the framework of SFA, the ionization amplitude can be evaluated by the SPA when the condition $\omega_0 \ll I_p$ is satisfied. Accordingly, one can introduce complex SPA-based trajectories that describe the under-the-barrier dynamics as tunneling with increasing energy (at $\gamma > 1$). In contrast, in  high-frequency fields $\omega_0 \sim I_p$, the SPA becomes invalid and the quantum orbit picture fails. In the present paper we study the time delay in a low frequency field  ($\omega_0 \ll I_p$) across the regimes from the quasistatic regime ($\gamma \ll 1$), nonadiabatic tunneling  ($\gamma \sim 1$) to the multiphoton regime ($\gamma \gg 1$).

The time delay is observed as a shift of the PMD peak with respect to the simple man model, when the electron appears in the continuum at the peak of the laser field and propagates afterward classically. The value of the time delay is evaluated using a  classical mapping of the ionization time to the asymptotic momentum.  

The structure of the paper is the following. In Sec.~\ref{sec:model} the SFA formalism is presented. The results  are presented in Sec.~\ref{sec:results}, and the role of the bound state depletion is discussed in Sec.~\ref{sec:discuss}. Our conclusions are given in Sec.~\ref{sec:concl}.

\section{The theoretical approach}\label{sec:model}

We consider a one-dimensional problem of strong field ionization. A single active electron is bound in a short-range atomic potential and is exposed to a half-cycle strong laser field. We solve the corresponding 
Schr\"odinger equation 
\begin{eqnarray}
i\partial_t|\psi(t)\rangle=H|\psi(t)\rangle,
\end{eqnarray}
with the Hamiltonian
\begin{eqnarray}
H=\hat{p}^2/2+V(x)+H^f(t),
\end{eqnarray}
where $\hat{p}$ is the momentum operator, $V(x)=-\kappa\delta(x)$ the atomic short-range potential, and $H^f(t)=xE(t)$ the laser-atom interaction Hamiltonian in the length gauge, with the laser electric field 
\begin{eqnarray}
E(t)=-E_0\exp[-(\omega_0 t)^2/2],
\end{eqnarray}
describing a half-cycle pulse. Initially the electron is in the bound state of the atomic potential 
\begin{eqnarray}
\psi^a_0(x,t)=\sqrt{\kappa}\exp(-\kappa\sqrt{x^2}+i\kappa^2t),
\end{eqnarray}
with the bound state energy $I_p=-\kappa^2/2$.

Our calculations are based on the SFA approach. For the description of the electron ionization dynamics, we begin with the expression of the exact solutions of TDSE via the Dyson equations  \cite{Becker_2002}:
\begin{eqnarray}
U(t,t')&=& U^a(t,t')-i\int_{t'}^t ds U(t,s) H^f(s) U^a(s,t')\label{U1}\\
U(t,t')&=&U^f(t,t')-i\int_{t'}^t ds U(t,s) V U^f(s,t'),\label{U2}
\end{eqnarray}
where $U(t,t')$ is the exact time-evolution operator (TEO) of the system, $U^f(t,t')$ the electron TEO in the laser field, and $U^a(t,t')$ the atomic TEO. Using Eq.~(\ref{U2}) in Eq.~(\ref{U1}), the second-order iterated  Dyson equation is obtained:
\begin{eqnarray}
U(t,t')&=& U^a(t,t')-i\int_{t'}^t ds U^f(t,s) H^f(s) U^a(s,t')\label{U3}\\
 &-&\int^{t}_{t'} dr\int^{r}_{t'}ds U(t,r )VU^f(r,s)H^f(s) U^a(s,t') .\nonumber
\end{eqnarray}
In the standard SFA, the exact TEO in the last term of Eq.~(\ref{U3}) is approximated by $U^f(t,t')$, i.e. by the first term of the representation of Eq.~(\ref{U2}), with the argumentation that the dynamics in the continuum is dominated by the laser field. 

In contrast, during the under-the-barrier recollision the effect of the atomic potential is important. Consequently, for the description of the under-the-barrier dynamics, we modify the standard SFA, using the representation of Eq.~(\ref{U1}) for the exact TEO in Eq.~(\ref{U3}). We include also the contribution of the second term of Eq.~(\ref{U1}) describing the rescattering, and in the last term approximate $U(t,t')\approx U^f(t,t')$, because after the rescattering the
laser field dominates the dynamics. In this way a modified high-order SFA approximation is derived:
\begin{eqnarray}
U(t,t')&\approx & U^a(t,t')-i\int_{t'}^t ds U^f(t,s) H^f(s) U^a(s,t')\label{U4}\\
 &-&\int^{t}_{t'} dr\int^{r}_{t'}ds U^a(t,r )VU^f(r,s)H^f(s) U^a(s,t') \nonumber\\
  &+&i\int^{t}_{t'} dr\int^r_{t'} ds\int^{s}_{t'}dt U^f(t,r )H^f(r) \nonumber \\
  &\times & U^a(r,s)VU^f(r,s)H^f(s) U^a(s,t'). \nonumber
\end{eqnarray}
With   the Volkov state $|\psi^f_p(t)\rangle$ of the electron in a laser field,  the Volkov TEO reads:
\begin{equation}
\label{fTEO}
U^f(t_f,r)= \int dq\,|\psi^f_q(t_f)\rangle\langle\psi^f_q(r)|,
\end{equation}
and the atomic TEO 
\begin{equation}
\label{aTEO}
U^a(t_f,r)=|\psi^a_0(t_f)\rangle\langle\psi^a_0(r)|+\int dq|\psi^a_q(t_f)\rangle\langle\psi^a_q(r)|,
\end{equation}
with the asymptotic scattering state $|\psi^a_p(t)\rangle$ of the atomic potential.

The solution of TDSE $|\psi(t_f)\rangle=U(t_f,t)|\psi^a_0(t)\rangle$ is approximated using the TEO of Eq.~(\ref{U4}), obtaining the  asymptotic momentum amplitude:
\begin{eqnarray}
m(p)=\langle\psi^a_p(t_f)|\psi(t_f)\rangle,
\end{eqnarray}
Accordingly, the PMD is calculated as
\begin{eqnarray}
\frac{dw}{dp}=|m(p)|^2.
\end{eqnarray}

The momentum amplitude contains two terms $m=m_1+m_{2}$, corresponding to the expansion of Eq.~(\ref{U4}). The first term $m_1$ is the well-known direct ionization amplitude:
 \begin{eqnarray}
\label{m1}
m_1(p)=-i\int dt\langle\psi^f_p(t)|H^f(t)|\psi^a_0(t)\rangle ,
\end{eqnarray}
describing the transition from the  bound state into the tunneling state. The second term describes the recollision-scattering and can be separated into two channels  $m_2=m_{2,\rm res}+m_{2,\rm rec}$, describing rescattering, and the recombination-ionization. The first channel amplitude is
\begin{eqnarray}
m_{2,\rm res}(p)&=&\int dr\int^{r}dt\int dq\langle\psi^a_p(r)|V|\psi^f_q(r)\rangle\langle\psi^f_q(t)|H^f(t)|\psi^a_0(t)\rangle\nonumber\\
&+&i\int dr\int^rds\int^sdt\int dq'\int dq\langle\psi^f_p(r)|H^f(r)|\psi^a_{q'}(r)\rangle\nonumber\\
&\times&\langle\psi^a_{q'}(s)|V|\psi^f_q(s)\rangle\langle\psi_q^f(t)|H^f(t)|\psi^a_0(t)\rangle,
\label{m2res}
\end{eqnarray}
which describes the rescattering channel (transition from the bound state into tunneling state with further scattering off the potential).
The recollision-recombination amplitude is:
\begin{eqnarray}
m_{2,\rm rec}(p)&=& i\int dr\int^{r}ds\int^sdt\int dq\langle\psi^f_p(r)|H^f(r)|\psi^a_0(r)\rangle\nonumber\\
&\times&\langle\psi^a_0(s)|V|\psi^f_q(s)\rangle\langle\psi_q^f(t)|H^f(t)|\psi^a_0(t)\rangle.
\label{m2rec}
\end{eqnarray}
describing the transition from  the bound state into tunneling state, reflection at the exit, rescattering at the  potential  into the bound state, with the follow up ionization into the continuum.
  
  \begin{figure*}
  \begin{center}
\includegraphics[width=0.7\textwidth]{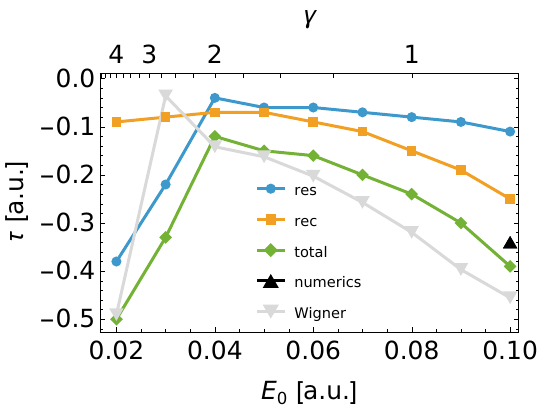}
 \caption{ Time delay in strong-field ionization  vs laser field strength: (blue line) the rescattering contribution, deduced from the PMD via $m_1(p)+m_{2,\rm res}(p)$ in Eqs.~(\ref{m1}), (\ref{m2res}), (orange line) the recombination contribution, deduced from the PMD via $m_1(p)+ m_{2,\rm rec}(p)$ in Eqs.~(\ref{m1}), (\ref{m2rec}), (green line)  the total time delay via the total PMD in Eqs.~(\ref{m1})-(\ref{m2rec}). The grey line shows the Wigner time delay  via Eq.~(\ref{tau}). The black triangle is the numerical result from the  Supplemental Material of Ref.~\cite{Klaiber_2022R}. }
\label{fig1}
\end{center}
\end{figure*}
 
The recollision term $m_2=m_{2,\rm res}+m_{2,\rm rec}$ in the applied modified SFA has several advantages with respect to the standard second-order SFA amplitude:
\begin{eqnarray}
\label{m2standard}
\tilde{m}_{2,\rm res}(p)&=&-\int dr\int^{r}dt\int dq\langle\psi^f_p(r)|V|\psi^f_q(r)\rangle\\
&\times&\langle\psi^f_q(t)|H^f(t)|\psi^a_0(t)\rangle\nonumber.
\end{eqnarray}
The standard SFA describes the rescattering in the Born approximation (transition between Vokov states) which is not a good approximation for the under-the barrier rescattering. In contrast,  in the applied SFA  here the scattered state is modified incorporating the atomic propagator. Moreover, our theory includes the depletion of the bound state during the interaction with the laser field, indicated by the matrix elements with $H^f(r)$ in the last step of the process [in the second term of Eq.~(\ref{m2res}); see also Sec.~\ref{sec:discuss} below]. 

When calculating the total ionization amplitude $m(p)=m_1(p)+ m_{2,\rm res}(p)+m_{2,\rm rec}(p)$, all time integrals are performed numerically, while the momentum integrals in the recombination and rescattering amplitudes are calculated analytically  using Green's functions:
\begin{widetext}
\begin{eqnarray}
G^a_0(x,s,y,t)&=&\psi^a_0(x,s)\psi^{a*}_0(y,t) = \kappa\exp[-\kappa(x+y)+i\kappa^2/2(s-t)]\nonumber \\
G^a(x,s,0,t)&=&\int dq\psi^a_q(x,s)\psi^{a*}_q(0,t)\approx\frac{\left(\frac{1}{2}+\frac{i}{2}\right) | x|  e^{\frac{i x^2}{2 s-2 t}}}{\sqrt{\pi } \kappa  (s-t)^{3/2}}\\
G^f(x,s,y,t)&=&\int dq\psi^f_q(x,s)\psi^{f*}_q(y,t)=\frac{\exp \left\{\frac{i}{2}  \left[\frac{(-\alpha
   (s)+\alpha (t)+x-y)^2}{s-t}+2 x \alpha '(s)-2 \beta (s)+2 \beta (t)-2 y
   \alpha '(t)\right]\right\}}{\sqrt{2 \pi i} \sqrt{s-t}}.\nonumber
\end{eqnarray}
\end{widetext}
The approximation $|q|\ll\kappa$ in the second equation is justified, since the under-the-barrier rescattering into bare scattering states can only happen close to the threshold of $q\sim 0$.
The remaining coordinate integrals are calculated analytically similar to Ref.~\cite{Klaiber_2023}, expressing them via the integral
\begin{eqnarray}
{\cal I}_i &=&\int^\infty_{-\infty}\, d \xi \; \xi\; \exp\left[ a_i\,\xi +b_i\, \xi^2 -\kappa \, \lvert \, \xi\, \rvert \right]\label{i11}\\
&=&\frac{\sqrt{\pi } e^{-\frac{\left( a_i + \kappa \right)^2}{4 b_i}}}{4 \left(-b_i\right)^{3/2}}\left\{e^{\frac{a_2 \kappa }{b_i}} \left(a_i-\kappa \right)
   \left[1+\text{erf}\left(\frac{a_i-\kappa }{2 \sqrt{-b_i}}\right)\right]\right.\nonumber\\
   &+&\left.\left(a_i+\kappa \right) \left(1-\text{erf}\left(\frac{a_i+\kappa }{2 \sqrt{-b_i}}\right) \right)\right\},\nonumber
\end{eqnarray}
with given coefficients $a_i$ and $b_i$.

Further, the boundaries of the time integrals are subtracted as in Ref.~\cite{Klaiber_2022edge,Klaiber_2026}. The boundaries of the finite time integration introduce not physically relevant time diffraction oscillations in the integration result. To avoid it, we shift the outer time integration contour in the complex plane:
\begin{eqnarray}
\int dt f(t)=f(t_i)^2/f'(t_i)+\int ^{t_f}_{t_i}dt f(t)-f(t_f)^2/f'(t_f),
\end{eqnarray} 
and omit the contribution of the vertical contours, which are responsible for the time diffraction.

  \begin{figure}[b]
  \begin{center}
  \includegraphics[width=0.45\textwidth]{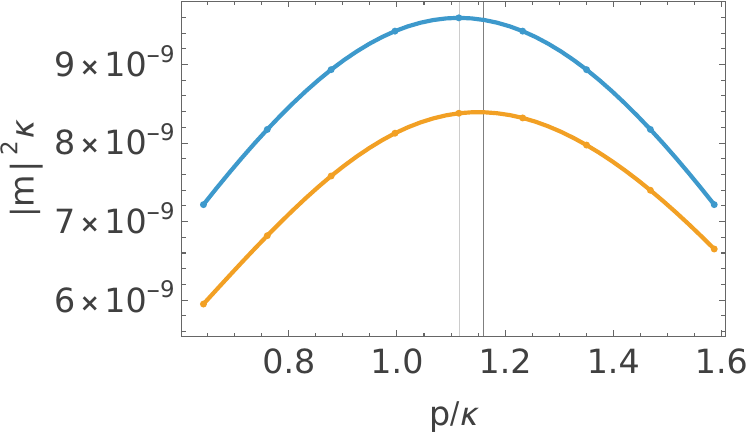}
 \caption{The PMD in the case of $E_0=0.02$ via $|m_1|^2$ (blue line) and the total $|m_1+m_2|^2$ (orange line). The peak of PMD is shifted ($\delta p$) due to the under-the-barrier recollision. The tunneling time is calculated as $\tau=-\delta p/E_0$.}
\label{fig22}
\end{center}
\end{figure}

\section{Results}\label{sec:results}

We have calculated PMDs and deduced the tunneling time delay at different field strengths, see results in Fig.~\ref{fig1}. The tunneling time delay is derived comparing the shift of the PMD peak $\delta p$ with respect to that via the direct ionization amplitude $|m_1|^2$, see Fig.~\ref{fig22}. Note that the PMD via the direct ionization amplitude coincides with the simple man picture, where the electron appears in the continuum most probably at the peak of the laser field ($t=t_0=0$) and propagates afterward classically, yielding the photoelectron momentum $p=-A(t_0)$, determined by the laser vector potential $A(t)$. The ionization time delay is calculated using the quasiclassical time-to-momentum mapping $\tau=-\delta p/E_0$ \cite{Klaiber_2022R}.

The approximate expression for the tunneling time can be estimated via the Wigner time delay as \cite{Klaiber_2024}:
\begin{eqnarray}
\label{tau}
\tau\approx -\frac{\partial}{\partial I_p}{\rm Im}\left\{ \frac{ m_{2,\rm res}+ m_{2,\rm rec}}{m_1}\right\} ,
\end{eqnarray}
which  shows a qualitatively correct behavior of the time delay, see the grey line in Fig.~\ref{fig1}, but a quantitatively correct result for the tunneling time is derived from the numerically calculated PMD.

The time delay dependence on the field strength is shown in Fig.~\ref{fig1}, along with the contributions of the rescattering in the PMD via $m_1(p)+m_{2,\rm res}(p)$, the recombination via $m_1(p)+m_{2,\rm rec}(p)$,  and the total time delay via the total PMD $m_1(p)+m_{2,\rm res}(p)+m_{2,\rm rec}(p)$. We use a laser angular frequency of $\omega_0= 0.08$ a.u., and an atomic system with $\kappa=1$ a.u.  and vary the laser field strength from $E_0=0.02$ ($\gamma\sim 4$) to $E_0=0.1$ a.u. ($\gamma\sim 0.8$), i.e. we span the interaction regime from  the nonadiabatic tunneling regime at large fields to the multiphoton regime at weak fields, and can distinguish between the influence of the under-the-barrier recombination and the under-the-barrier rescattering on the tunneling time.

For large fields in the quasistatic regime the recombination dominates and yields  a negative time delay due to interference with the direct trajectory.  There is almost no energy gain during tunneling and the recollision energy is largely negative ($\sim -I_p$). The scattering into atomic eigenstates is therefore only possible into the ground state. In Ref.~\cite{Klaiber_2018} the recombination contribution was included via the use of the exact laser-free scattering $T$-matrix, as in the quasistatic regime the low-frequency approximation \cite{Cerkic_2009} is valid. 

For weak fields in the nonadiabatic regime the rescattering has the largest impact since the energy gain is now large and recombination into the scattering atomic states becomes possible, whereas recombination into the bound state is unlikely.

  \begin{figure}
  \begin{center}
  \includegraphics[width=0.4\textwidth]{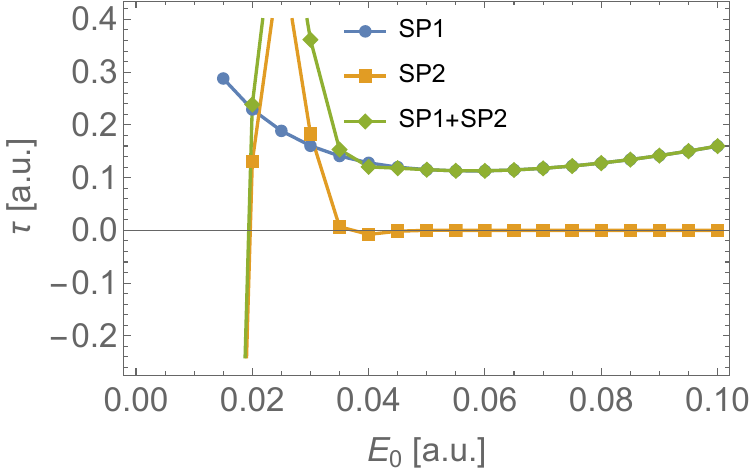}
\includegraphics[width=0.4\textwidth]{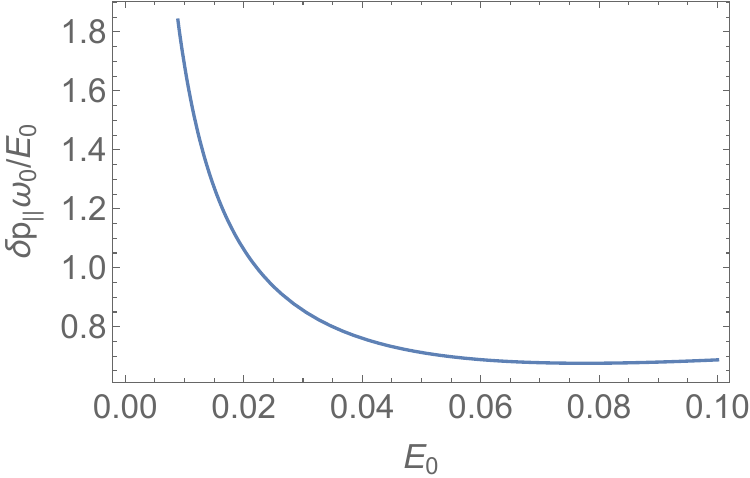}
 \caption{Tunneling time in Born approximation via a numerical integration and via the saddle point approximation: (a) with first saddle-point of the ordinary trajectory (1,1,0) (blue), with second  saddle point of the extraordinary trajectory (-1,1,0) (orange) and  ordinary, and extraordinary one together (green), (b) Ratio of the longitudinal momentum width $\delta p_\Vert$ to the available momentum in the laser field. }
\label{fig2}
\end{center}
\end{figure}

To confirm the steep increase of the module of the negative time delay at $E_0<0.04$ due to rescattering, we carried out a rough estimation applying SPA. To this end the use of Eq.~(\ref{m2rec}) is complicated because of three time-integrations. To enable SPA we employ the standard SFA scheme of Eq.~(\ref{m2standard}), which is physically less accurate than the employed SFA Eq.~(\ref{U4}), but amenable to SPA calculations. The results are presented in Fig.~\ref{fig2}(a). To have a good coincidence of the SPA with the numerical result, besides the ordinary saddle-point (SP) for the under-the-barrier trajectory (the trajectory (1,1,0) in the Milo\v{s}evi\'c's classification~\cite{Milosevic_2014}), one needs to include the second extraordinary quantum orbit related to the under-the-barrier motion, the trajectory (-1,1,0), visualized in Fig.~9 of \cite{Milosevic_2014}. The trajectory (-1,1,0)  starts at the end of half-cycle pulse, where the canonical (drift) momentum is  larger by modulus than $E_0/\omega_0$, and negative. It is only relevant when the longitudinal momentum width  $\delta p_\Vert=\sqrt{\omega_0 [\sinh^{-1}(\gamma) -\gamma/\sqrt{1+\gamma^2} ]} $~\cite{Popov_2004u} is larger than $E_0/\omega_0$, which is only the case in the nonadiabatic regime with $\gamma\gtrsim 1$. For this trajectory the initial momentum at the tunnel exit is negative, which automatically involves sub-barrier dynamics. Due to this negative momentum a recollision even in a half-cycle pulse is possible. 
From Fig.~\ref{fig2}(b)  one can see that $\delta p_\Vert > E_0/\omega_0$ for the field strengths of approximately $E_0<0.025$ a.u. and the extraordinary trajectory contributes below this field value as Fig.~\ref{fig2}(a) also illustrates. Thus, the large negative tunneling time at weak fields is due to the effect of this extraordinary under-the-barrier trajectory.

\section{The bound state distortion and depletion}\label{sec:discuss}
\begin{figure}[b]
  \begin{center}
\includegraphics[width=0.45\textwidth]{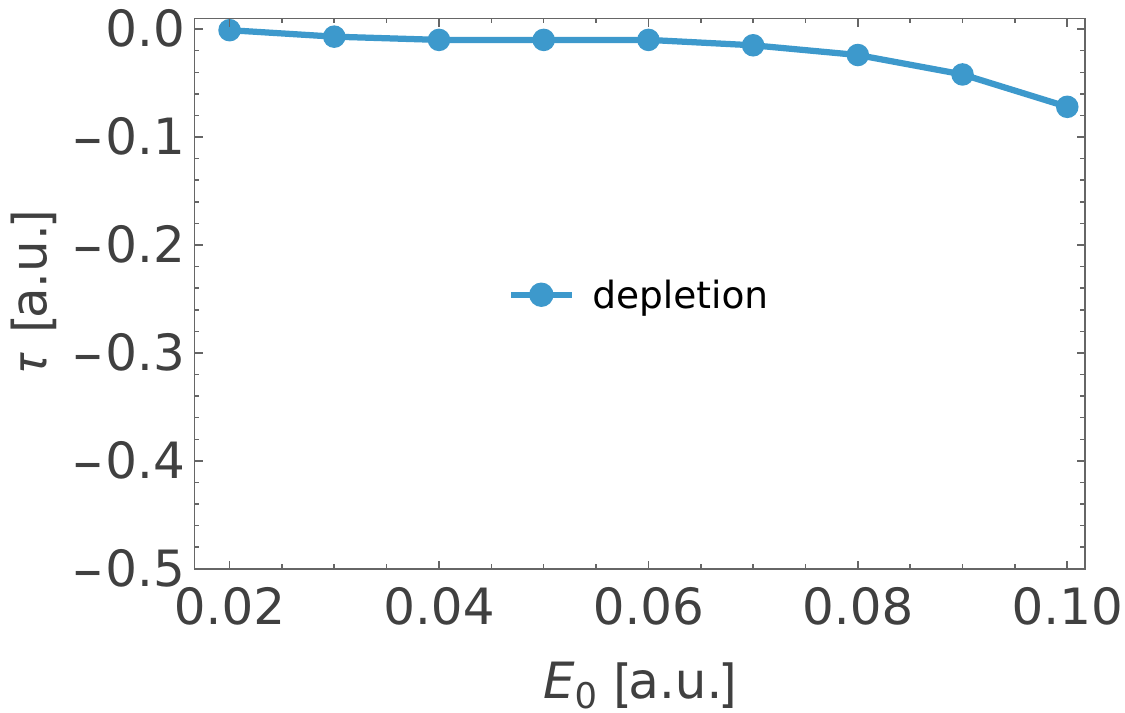}
 \caption{Depletion contribution to the tunneling time:  $\tau=-\delta p/E_0$, with $\delta p$ determined from the shift of the peak of the PMD  $m_1+m_{2,\rm depl}$ with respect to that of $m_1$.}
\label{fig4}
\end{center}
\end{figure}

In this paper we intentionally do not apply SPA for calculation of integrals, because the numerical calculation of the full integrals includes the contribution of the bound state distortion in the laser field. Our theory includes the distortion of the bound state during the interaction with the laser field. It can be seen by rewriting the second order rescattering term   by a partial integration from $H^f$ to $V$ matrix element \cite{Lohr_1997}:
\begin{eqnarray}
\label{m2V}
m_2(p)&=&-\int^{t_f} dr\int^{r}dt  \langle\psi^f_p(r)|V|U^f(r,t)|V|\psi^a_0(t)\rangle.
\end{eqnarray}
It contains  the direct path and the rescattering path via Born approximation. When applying SPA, the contour in the $t$-integral goes from $-\infty$ to the saddle-point $t_s$ horizontally to the real $t$-axis and describes the rescattering, whereas the second, vertical  part of the contour from the saddle-point $t_s$ to that of $r_s$ does the direct ionization \cite{Krajewska_2010}. In fact, the vertical part 
\begin{eqnarray}
\label{m2V}
m^v_2(p)&=&-\int^{t_f} dr\int^{r}_{t_s}dt \langle\psi^f_p(r)|V|U^f(r,t)|V|\psi^a_0(t)\rangle.
\end{eqnarray}
can be expanded around $t=r_s$ and yields in the quasistatic approximation 
\begin{eqnarray}
m^v_2(p)&=&m_1[1+5/8(E_0/E_a^2],
\end{eqnarray}
which coincides with the distortion of the bound state at the core \cite{Krajewska_2010,Tolstikhin_2013}. 

We note also that the recombination amplitude contains the depletion effect of the atomic bound state.
This can be seen when applying a partial integration of the middle matrix element in Eq.~(\ref{m2rec}) of the recombination amplitude $m_{2,\rm rec}$, transforming it from a $V$-form $\langle\psi^a_0(s)|V|\psi^f_q(s)\rangle$ to a $H^f$-form $\langle\psi^a_0(s)|H^f|\psi^f_q(s)\rangle$, and analyzing the nested time integrals. The latter can be expressed as follows, see Appendix \ref{rec}:
\begin{eqnarray}
\int^rds\int^sdt\,m^*(s)\,m(t)&=& i{\rm Im}\left\{\int^rdt\int^r_tdsm^*(s)\,m(t)\right\}\nonumber\\
&+&\frac{1}{2}\int^rds\int^rdtm^*(s)\,m(t), 
\end{eqnarray}
where  $m$ stands for the matrix element in the $H^f$-form. The real part of the recombination amplitude is therefore the depletion effect $m_{\rm depl}\propto\frac{1}{2}\int^rds\int^rdtm^*(s)\,m(t) $, whereas the imaginary part describes the effect of the recombination into the bound state after tunneling. Thus, the depletion  is defined via 
\begin{eqnarray}
m_{2,\rm depl}(p)&=&  i\int dr\langle\psi^f_p(r)|H^f(r)|\psi^a_0(r)\rangle\int^{r}ds\int^rdt\int dq\nonumber\\
&\times& \langle\psi^a_0(s)|H^f|\psi^f_q(s)\rangle\langle\psi_q^f(t)|H^f(t)|\psi^a_0(t)\rangle/2.
\label{m2rec1}
\end{eqnarray}
This allows for the estimation of the contribution of the depletion effect to the tunneling time, see Fig.~\ref{fig4}.  The contribution of the depletion is included in the time delay which is shown in Fig.~\ref{fig1}, but Fig.~\ref{fig4} demonstrates that the depletion contribution is small and especially negligible in the nonadiabatic regime.

\section{Conclusion}\label{sec:concl}

We have calculated the tunneling time delay systematically within the SFA formalism with  high-order corrections. The theory is applied for the tunneling ionization of a one-dimensional atom with a short-range potential in a half-cycle pulse. In the applied modified SFA scheme, the second-order  amplitude describes not only the under-the-barrier rescattering, but also  a new channel, which can be interpreted as  recombination during the recollision with a further release from the bound state. This new channel was implicitly included in the SFA analysis of the tunneling time in the quasistatic regime in Ref.~\cite{Klaiber_2018} via employing heuristically the exact scattering $T$-matrix in the SFA amplitude.
However, in the nonadiabatic regime the use of the low-frequency approximation and the field-free  $T$-matrix in SFA is not valid. Therefore, for the  treatment of the tunneling time in the nonadiabatic regime, we had to use a modified high-order SFA, specifically tailored for the under-the-barrier dynamics by combining different versions of the Dyson equation for the exact TEO, and approximating the exact TEO in the third iteration. We use SFA calculations beyond SPA, which allows us to derive accurately  the behavior of the tunneling time delay with respect to the laser field. We demonstrated a significant enhancement of the negative time delay in the   multiphoton ionization regime abruptly at $\gamma<2$ in weak fields. 

For the large fields $\gamma\ll 1$ (for the chosen parameters at $E_0\gtrsim 0.1$), the recombination is the main channel to generate the negative tunneling time delay.  In weak fields when the nonadiabatic tunneling regime is transferred to the multiphoton regime $\gamma\gg 1$ ($E_0\lesssim 0.04$), the significant increase of the module of the negative ionization time delay is due to the under-the-barrier rescattering. In the language of the SPA, we can explain  this behavior by the contribution of the extraordinary saddle-point related to the extraordinary under-the barrier trajectory at weak fields. 

Although derived from a one-dimensional framework, these findings retain qualitative validity for strong-field ionization in circularly polarized laser fields, effectively characterizing the absolute PMD peak shift in the nonadiabatic regime.\\

\appendix

\section{Manipulation of the recombination amplitude}\label{rec}

The nested time integrals  in Eq.~(\ref{m2res}), (\ref{m2rec}) can be brought into $H^f$-form and then manipulated in the following way:
 \begin{eqnarray}
&&\int^rds\int^sdtm^*(s)m(t)=\int^rdt\int^r_tdsm^*(s)m(t)\\
&&=\int^rds\int^sdtm^*(s)m(t)/2+\int^rds\int^sdtm^*(s)m(t)/2\nonumber\\
&&-\int^rds\int^rdtm^*(s)m(t)/2+\int^rds\int^rdtm^*(s)m(t)/2\nonumber\\
&&=\int^rds\int^sdtm^*(s)m(t)/2+\int^rds\int^s_rdtm^*(s)m(t)/2\nonumber\\
&&+\int^rds\int^rdtm^*(s)m(t)/2\nonumber\\
&&=\int^rds\int^sdtm^*(s)m(t)/2-\left(\int^rdt\int^r_tdsm^*(s)m(t)\right)^*/2\nonumber\\
&&+\int^rds\int^rdtm^*(s)m(t)/2\nonumber\\
&&=i{\rm Im}\left\{\int^rdt\int^r_tdsm^*(s)m(t)\right\}+\int^rds\int^rdtm^*(s)m(t)/2\nonumber.
\end{eqnarray}

\bibliography{strong_fields_bibliography}

\begin{thebibliography}{52}%
\makeatletter
\providecommand \@ifxundefined [1]{%
 \@ifx{#1\undefined}
}%
\providecommand \@ifnum [1]{%
 \ifnum #1\expandafter \@firstoftwo
 \else \expandafter \@secondoftwo
 \fi
}%
\providecommand \@ifx [1]{%
 \ifx #1\expandafter \@firstoftwo
 \else \expandafter \@secondoftwo
 \fi
}%
\providecommand \natexlab [1]{#1}%
\providecommand \enquote  [1]{``#1''}%
\providecommand \bibnamefont  [1]{#1}%
\providecommand \bibfnamefont [1]{#1}%
\providecommand \citenamefont [1]{#1}%
\providecommand \href@noop [0]{\@secondoftwo}%
\providecommand \href [0]{\begingroup \@sanitize@url \@href}%
\providecommand \@href[1]{\@@startlink{#1}\@@href}%
\providecommand \@@href[1]{\endgroup#1\@@endlink}%
\providecommand \@sanitize@url [0]{\catcode `\\12\catcode `\$12\catcode
  `\&12\catcode `\#12\catcode `\^12\catcode `\_12\catcode `\%12\relax}%
\providecommand \@@startlink[1]{}%
\providecommand \@@endlink[0]{}%
\providecommand \url  [0]{\begingroup\@sanitize@url \@url }%
\providecommand \@url [1]{\endgroup\@href {#1}{\urlprefix }}%
\providecommand \urlprefix  [0]{URL }%
\providecommand \Eprint [0]{\href }%
\providecommand \doibase [0]{https://doi.org/}%
\providecommand \selectlanguage [0]{\@gobble}%
\providecommand \bibinfo  [0]{\@secondoftwo}%
\providecommand \bibfield  [0]{\@secondoftwo}%
\providecommand \translation [1]{[#1]}%
\providecommand \BibitemOpen [0]{}%
\providecommand \bibitemStop [0]{}%
\providecommand \bibitemNoStop [0]{.\EOS\space}%
\providecommand \EOS [0]{\spacefactor3000\relax}%
\providecommand \BibitemShut  [1]{\csname bibitem#1\endcsname}%
\let\auto@bib@innerbib\@empty
\bibitem [{\citenamefont {Popov}(2005)}]{Popov_2005}%
  \BibitemOpen
  \bibfield  {author} {\bibinfo {author} {\bibfnamefont {V.~S.}\ \bibnamefont
  {Popov}},\ }\bibfield  {title} {\bibinfo {title} {Imaginary-time method in
  quantum mechanics and field theory},\ }\href@noop {} {\bibfield  {journal}
  {\bibinfo  {journal} {Phys. Atom. Nuclei}\ }\textbf {\bibinfo {volume}
  {68}},\ \bibinfo {pages} {686} (\bibinfo {year} {2005})}\BibitemShut
  {NoStop}%
\bibitem [{\citenamefont {Perelomov}\ and\ \citenamefont
  {Popov}(1966)}]{Perelomov_1966a}%
  \BibitemOpen
  \bibfield  {author} {\bibinfo {author} {\bibfnamefont {A.~M.}\ \bibnamefont
  {Perelomov}}\ and\ \bibinfo {author} {\bibfnamefont {V.~S.}\ \bibnamefont
  {Popov}},\ }\bibfield  {title} {\bibinfo {title} {Ionization of atoms in an
  alternating electric field},\ }\href@noop {} {\bibfield  {journal} {\bibinfo
  {journal} {Zh. Exp. Theor. Fiz.}\ }\textbf {\bibinfo {volume} {50}},\
  \bibinfo {pages} {1393} (\bibinfo {year} {1966})}\BibitemShut {NoStop}%
\bibitem [{\citenamefont {Perelomov}\ \emph {et~al.}(1966)\citenamefont
  {Perelomov}, \citenamefont {Popov},\ and\ \citenamefont
  {Terent'ev}}]{Perelomov_1967a}%
  \BibitemOpen
  \bibfield  {author} {\bibinfo {author} {\bibfnamefont {A.~M.}\ \bibnamefont
  {Perelomov}}, \bibinfo {author} {\bibfnamefont {V.~S.}\ \bibnamefont
  {Popov}},\ and\ \bibinfo {author} {\bibfnamefont {V.~M.}\ \bibnamefont
  {Terent'ev}},\ }\bibfield  {title} {\bibinfo {title} {Ionization of atoms in
  an alternating electric field. ii},\ }\href@noop {} {\bibfield  {journal}
  {\bibinfo  {journal} {Zh. Exp. Theor. Fiz.}\ }\textbf {\bibinfo {volume}
  {51}},\ \bibinfo {pages} {309} (\bibinfo {year} {1966})}\BibitemShut
  {NoStop}%
\bibitem [{\citenamefont {Perelomov}\ and\ \citenamefont
  {Popov}(1967)}]{Perelomov_1967b}%
  \BibitemOpen
  \bibfield  {author} {\bibinfo {author} {\bibfnamefont {A.~M.}\ \bibnamefont
  {Perelomov}}\ and\ \bibinfo {author} {\bibfnamefont {V.~S.}\ \bibnamefont
  {Popov}},\ }\bibfield  {title} {\bibinfo {title} {Ionization of atoms in an
  alternating electric field. iii},\ }\href@noop {} {\bibfield  {journal}
  {\bibinfo  {journal} {Zh. Exp. Theor. Fiz.}\ }\textbf {\bibinfo {volume}
  {52}},\ \bibinfo {pages} {514} (\bibinfo {year} {1967})}\BibitemShut
  {NoStop}%
\bibitem [{\citenamefont {Popov}(2004)}]{Popov_2004u}%
  \BibitemOpen
  \bibfield  {author} {\bibinfo {author} {\bibfnamefont {V.~S.}\ \bibnamefont
  {Popov}},\ }\bibfield  {title} {\bibinfo {title} {{Tunneling and multiphoton
  ionization of atoms and ions ina strong laser field (Keldysh theory)}},\
  }\href@noop {} {\bibfield  {journal} {\bibinfo  {journal} {Phys. Usp.}\
  }\textbf {\bibinfo {volume} {47}},\ \bibinfo {pages} {855} (\bibinfo {year}
  {2004})}\BibitemShut {NoStop}%
\bibitem [{\citenamefont {Keldysh}(1964)}]{Keldysh_1965}%
  \BibitemOpen
  \bibfield  {author} {\bibinfo {author} {\bibfnamefont {L.~V.}\ \bibnamefont
  {Keldysh}},\ }\bibfield  {title} {\bibinfo {title} {Ionization in the field
  of a strong electromagnetic wave},\ }\href@noop {} {\bibfield  {journal}
  {\bibinfo  {journal} {Zh. Eksp. Teor. Fiz.}\ }\textbf {\bibinfo {volume}
  {47}},\ \bibinfo {pages} {1945} (\bibinfo {year} {1964})},\ \bibinfo {note}
  {[Sov. Phys. JETP 20, 1307 (1965)]}\BibitemShut {NoStop}%
\bibitem [{\citenamefont {Yan}\ and\ \citenamefont {Bauer}(2012)}]{Yan_2012a}%
  \BibitemOpen
  \bibfield  {author} {\bibinfo {author} {\bibfnamefont {T.-M.}\ \bibnamefont
  {Yan}}\ and\ \bibinfo {author} {\bibfnamefont {D.}~\bibnamefont {Bauer}},\
  }\bibfield  {title} {\bibinfo {title} {Sub-barrier coulomb effects on the
  interference pattern in tunneling-ionization photoelectron spectra},\
  }\href@noop {} {\bibfield  {journal} {\bibinfo  {journal} {Phys. Rev. A}\
  }\textbf {\bibinfo {volume} {86}},\ \bibinfo {pages} {053403} (\bibinfo
  {year} {2012})}\BibitemShut {NoStop}%
\bibitem [{\citenamefont {Huismans}\ \emph {et~al.}(2011)\citenamefont
  {Huismans}, \citenamefont {Rouz{\'{e}}e}, \citenamefont {Gijsbertsen},
  \citenamefont {Jungmann}, \citenamefont {Smolkowska}, \citenamefont {Logman},
  \citenamefont {L{\'{e}}pine}, \citenamefont {Cauchy}, \citenamefont {Zamith},
  \citenamefont {Marchenko}, \citenamefont {Bakker}, \citenamefont {Berden},
  \citenamefont {Redlich}, \citenamefont {van~der Meer}, \citenamefont
  {Muller}, \citenamefont {Vermin}, \citenamefont {Schafer}, \citenamefont
  {Spanner}, \citenamefont {Ivanov}, \citenamefont {Smirnova}, \citenamefont
  {Bauer}, \citenamefont {Popruzhenko},\ and\ \citenamefont
  {Vrakking}}]{Huismans_2011}%
  \BibitemOpen
  \bibfield  {author} {\bibinfo {author} {\bibfnamefont {Y.}~\bibnamefont
  {Huismans}}, \bibinfo {author} {\bibfnamefont {A.}~\bibnamefont
  {Rouz{\'{e}}e}}, \bibinfo {author} {\bibfnamefont {A.}~\bibnamefont
  {Gijsbertsen}}, \bibinfo {author} {\bibfnamefont {J.~H.}\ \bibnamefont
  {Jungmann}}, \bibinfo {author} {\bibfnamefont {A.~S.}\ \bibnamefont
  {Smolkowska}}, \bibinfo {author} {\bibfnamefont {P.~S. W.~M.}\ \bibnamefont
  {Logman}}, \bibinfo {author} {\bibfnamefont {F.}~\bibnamefont
  {L{\'{e}}pine}}, \bibinfo {author} {\bibfnamefont {C.}~\bibnamefont
  {Cauchy}}, \bibinfo {author} {\bibfnamefont {S.}~\bibnamefont {Zamith}},
  \bibinfo {author} {\bibfnamefont {T.}~\bibnamefont {Marchenko}}, \bibinfo
  {author} {\bibfnamefont {J.~M.}\ \bibnamefont {Bakker}}, \bibinfo {author}
  {\bibfnamefont {G.}~\bibnamefont {Berden}}, \bibinfo {author} {\bibfnamefont
  {B.}~\bibnamefont {Redlich}}, \bibinfo {author} {\bibfnamefont {A.~F.~G.}\
  \bibnamefont {van~der Meer}}, \bibinfo {author} {\bibfnamefont {H.~G.}\
  \bibnamefont {Muller}}, \bibinfo {author} {\bibfnamefont {W.}~\bibnamefont
  {Vermin}}, \bibinfo {author} {\bibfnamefont {K.~J.}\ \bibnamefont {Schafer}},
  \bibinfo {author} {\bibfnamefont {M.}~\bibnamefont {Spanner}}, \bibinfo
  {author} {\bibfnamefont {M.~Y.}\ \bibnamefont {Ivanov}}, \bibinfo {author}
  {\bibfnamefont {O.}~\bibnamefont {Smirnova}}, \bibinfo {author}
  {\bibfnamefont {D.}~\bibnamefont {Bauer}}, \bibinfo {author} {\bibfnamefont
  {S.~V.}\ \bibnamefont {Popruzhenko}},\ and\ \bibinfo {author} {\bibfnamefont
  {M.~J.~J.}\ \bibnamefont {Vrakking}},\ }\bibfield  {title} {\bibinfo {title}
  {{Time-resolved holography with photoelectrons}},\ }\href
  {https://doi.org/10.1126/science.1198450} {\bibfield  {journal} {\bibinfo
  {journal} {Science (New York, N.Y.)}\ }\textbf {\bibinfo {volume} {331}},\
  \bibinfo {pages} {61} (\bibinfo {year} {2011})}\BibitemShut {NoStop}%
\bibitem [{\citenamefont {Keil}\ \emph {et~al.}(2016)\citenamefont {Keil},
  \citenamefont {Popruzhenko},\ and\ \citenamefont {Bauer}}]{Keil_2016}%
  \BibitemOpen
  \bibfield  {author} {\bibinfo {author} {\bibfnamefont {T.}~\bibnamefont
  {Keil}}, \bibinfo {author} {\bibfnamefont {S.~V.}\ \bibnamefont
  {Popruzhenko}},\ and\ \bibinfo {author} {\bibfnamefont {D.}~\bibnamefont
  {Bauer}},\ }\bibfield  {title} {\bibinfo {title} {{Laser-Driven Recollisions
  under the Coulomb Barrier}},\ }\href@noop {} {\bibfield  {journal} {\bibinfo
  {journal} {Phys. Rev. Lett.}\ }\textbf {\bibinfo {volume} {117}},\ \bibinfo
  {pages} {243003} (\bibinfo {year} {2016})}\BibitemShut {NoStop}%
\bibitem [{\citenamefont {He}\ \emph {et~al.}(2022)\citenamefont {He},
  \citenamefont {Klaiber}, \citenamefont {Hatsagortsyan},\ and\ \citenamefont
  {Keitel}}]{He_2022}%
  \BibitemOpen
  \bibfield  {author} {\bibinfo {author} {\bibfnamefont {P.-L.}\ \bibnamefont
  {He}}, \bibinfo {author} {\bibfnamefont {M.}~\bibnamefont {Klaiber}},
  \bibinfo {author} {\bibfnamefont {K.~Z.}\ \bibnamefont {Hatsagortsyan}},\
  and\ \bibinfo {author} {\bibfnamefont {C.~H.}\ \bibnamefont {Keitel}},\
  }\bibfield  {title} {\bibinfo {title} {Nondipole coulomb sub-barrier
  ionization dynamics and photon momentum sharing},\ }\href
  {https://doi.org/10.1103/PhysRevA.105.L031102} {\bibfield  {journal}
  {\bibinfo  {journal} {Phys. Rev. A}\ }\textbf {\bibinfo {volume} {105}},\
  \bibinfo {pages} {L031102} (\bibinfo {year} {2022})}\BibitemShut {NoStop}%
\bibitem [{\citenamefont {Klaiber}\ \emph {et~al.}(2018)\citenamefont
  {Klaiber}, \citenamefont {Hatsagortsyan},\ and\ \citenamefont
  {Keitel}}]{Klaiber_2018}%
  \BibitemOpen
  \bibfield  {author} {\bibinfo {author} {\bibfnamefont {M.}~\bibnamefont
  {Klaiber}}, \bibinfo {author} {\bibfnamefont {K.~Z.}\ \bibnamefont
  {Hatsagortsyan}},\ and\ \bibinfo {author} {\bibfnamefont {C.~H.}\
  \bibnamefont {Keitel}},\ }\bibfield  {title} {\bibinfo {title}
  {Under-the-tunneling-barrier recollisions in strong-field ionization},\
  }\href@noop {} {\bibfield  {journal} {\bibinfo  {journal} {Phys. Rev. Lett.}\
  }\textbf {\bibinfo {volume} {120}},\ \bibinfo {pages} {013201} (\bibinfo
  {year} {2018})}\BibitemShut {NoStop}%
\bibitem [{\citenamefont {Eckle}\ \emph
  {et~al.}(2008{\natexlab{a}})\citenamefont {Eckle}, \citenamefont {Smolarski},
  \citenamefont {Schlup}, \citenamefont {Biegert}, \citenamefont {Staudte},
  \citenamefont {Sch\"offler}, \citenamefont {Muller}, \citenamefont
  {D\"orner},\ and\ \citenamefont {Keller}}]{Eckle_2008a}%
  \BibitemOpen
  \bibfield  {author} {\bibinfo {author} {\bibfnamefont {P.}~\bibnamefont
  {Eckle}}, \bibinfo {author} {\bibfnamefont {M.}~\bibnamefont {Smolarski}},
  \bibinfo {author} {\bibfnamefont {F.}~\bibnamefont {Schlup}}, \bibinfo
  {author} {\bibfnamefont {J.}~\bibnamefont {Biegert}}, \bibinfo {author}
  {\bibfnamefont {A.}~\bibnamefont {Staudte}}, \bibinfo {author} {\bibfnamefont
  {M.}~\bibnamefont {Sch\"offler}}, \bibinfo {author} {\bibfnamefont {H.~G.}\
  \bibnamefont {Muller}}, \bibinfo {author} {\bibfnamefont {R.}~\bibnamefont
  {D\"orner}},\ and\ \bibinfo {author} {\bibfnamefont {U.}~\bibnamefont
  {Keller}},\ }\bibfield  {title} {\bibinfo {title} {Attosecond angular
  streaking},\ }\href@noop {} {\bibfield  {journal} {\bibinfo  {journal}
  {Nature Phys.}\ }\textbf {\bibinfo {volume} {4}},\ \bibinfo {pages} {565}
  (\bibinfo {year} {2008}{\natexlab{a}})}\BibitemShut {NoStop}%
\bibitem [{\citenamefont {Eckle}\ \emph
  {et~al.}(2008{\natexlab{b}})\citenamefont {Eckle}, \citenamefont {Pfeiffer},
  \citenamefont {Cirelli}, \citenamefont {Staudte}, \citenamefont {D\"orner},
  \citenamefont {Muller}, \citenamefont {B\"uttiker},\ and\ \citenamefont
  {Keller}}]{Eckle_2008b}%
  \BibitemOpen
  \bibfield  {author} {\bibinfo {author} {\bibfnamefont {P.}~\bibnamefont
  {Eckle}}, \bibinfo {author} {\bibfnamefont {A.~N.}\ \bibnamefont {Pfeiffer}},
  \bibinfo {author} {\bibfnamefont {C.}~\bibnamefont {Cirelli}}, \bibinfo
  {author} {\bibfnamefont {A.}~\bibnamefont {Staudte}}, \bibinfo {author}
  {\bibfnamefont {R.}~\bibnamefont {D\"orner}}, \bibinfo {author}
  {\bibfnamefont {H.~G.}\ \bibnamefont {Muller}}, \bibinfo {author}
  {\bibfnamefont {M.}~\bibnamefont {B\"uttiker}},\ and\ \bibinfo {author}
  {\bibfnamefont {U.}~\bibnamefont {Keller}},\ }\bibfield  {title} {\bibinfo
  {title} {Attosecond ionization and tunneling delay time measurements in
  helium},\ }\href@noop {} {\bibfield  {journal} {\bibinfo  {journal}
  {Science}\ }\textbf {\bibinfo {volume} {322}},\ \bibinfo {pages} {1525}
  (\bibinfo {year} {2008}{\natexlab{b}})}\BibitemShut {NoStop}%
\bibitem [{\citenamefont {Pfeiffer}\ \emph {et~al.}(2012)\citenamefont
  {Pfeiffer}, \citenamefont {Cirelli}, \citenamefont {Smolarski}, \citenamefont
  {Dimitrovski}, \citenamefont {Abu-samha}, \citenamefont {Madsen},\ and\
  \citenamefont {Keller}}]{Pfeiffer_2012}%
  \BibitemOpen
  \bibfield  {author} {\bibinfo {author} {\bibfnamefont {A.~N.}\ \bibnamefont
  {Pfeiffer}}, \bibinfo {author} {\bibfnamefont {C.}~\bibnamefont {Cirelli}},
  \bibinfo {author} {\bibfnamefont {M.}~\bibnamefont {Smolarski}}, \bibinfo
  {author} {\bibfnamefont {D.}~\bibnamefont {Dimitrovski}}, \bibinfo {author}
  {\bibfnamefont {M.}~\bibnamefont {Abu-samha}}, \bibinfo {author}
  {\bibfnamefont {L.~B.}\ \bibnamefont {Madsen}},\ and\ \bibinfo {author}
  {\bibfnamefont {U.}~\bibnamefont {Keller}},\ }\bibfield  {title} {\bibinfo
  {title} {Attoclock reveals natural coordinates of the laser-induced
  tunnelling current flow in atoms},\ }\href@noop {} {\bibfield  {journal}
  {\bibinfo  {journal} {Nature Phys.}\ }\textbf {\bibinfo {volume} {8}},\
  \bibinfo {pages} {76} (\bibinfo {year} {2012})}\BibitemShut {NoStop}%
\bibitem [{\citenamefont {Landsman}\ \emph {et~al.}(2014)\citenamefont
  {Landsman}, \citenamefont {Weger}, \citenamefont {Maurer}, \citenamefont
  {Boge}, \citenamefont {Ludwig}, \citenamefont {Heuser}, \citenamefont
  {Cirelli}, \citenamefont {Gallmann},\ and\ \citenamefont
  {Keller}}]{Landsman_2014o}%
  \BibitemOpen
  \bibfield  {author} {\bibinfo {author} {\bibfnamefont {A.~S.}\ \bibnamefont
  {Landsman}}, \bibinfo {author} {\bibfnamefont {M.}~\bibnamefont {Weger}},
  \bibinfo {author} {\bibfnamefont {J.}~\bibnamefont {Maurer}}, \bibinfo
  {author} {\bibfnamefont {R.}~\bibnamefont {Boge}}, \bibinfo {author}
  {\bibfnamefont {A.}~\bibnamefont {Ludwig}}, \bibinfo {author} {\bibfnamefont
  {S.}~\bibnamefont {Heuser}}, \bibinfo {author} {\bibfnamefont
  {C.}~\bibnamefont {Cirelli}}, \bibinfo {author} {\bibfnamefont
  {L.}~\bibnamefont {Gallmann}},\ and\ \bibinfo {author} {\bibfnamefont
  {U.}~\bibnamefont {Keller}},\ }\bibfield  {title} {\bibinfo {title}
  {Ultrafast resolution of tunneling delay time},\ }\href@noop {} {\bibfield
  {journal} {\bibinfo  {journal} {Optica}\ }\textbf {\bibinfo {volume} {1}},\
  \bibinfo {pages} {343} (\bibinfo {year} {2014})}\BibitemShut {NoStop}%
\bibitem [{\citenamefont {Camus}\ \emph {et~al.}(2017)\citenamefont {Camus},
  \citenamefont {Yakaboylu}, \citenamefont {Fechner}, \citenamefont {Klaiber},
  \citenamefont {Laux}, \citenamefont {Mi}, \citenamefont {Hatsagortsyan},
  \citenamefont {Pfeifer}, \citenamefont {Keitel},\ and\ \citenamefont
  {Moshammer}}]{Camus_2017}%
  \BibitemOpen
  \bibfield  {author} {\bibinfo {author} {\bibfnamefont {N.}~\bibnamefont
  {Camus}}, \bibinfo {author} {\bibfnamefont {E.}~\bibnamefont {Yakaboylu}},
  \bibinfo {author} {\bibfnamefont {L.}~\bibnamefont {Fechner}}, \bibinfo
  {author} {\bibfnamefont {M.}~\bibnamefont {Klaiber}}, \bibinfo {author}
  {\bibfnamefont {M.}~\bibnamefont {Laux}}, \bibinfo {author} {\bibfnamefont
  {Y.}~\bibnamefont {Mi}}, \bibinfo {author} {\bibfnamefont {K.~Z.}\
  \bibnamefont {Hatsagortsyan}}, \bibinfo {author} {\bibfnamefont
  {T.}~\bibnamefont {Pfeifer}}, \bibinfo {author} {\bibfnamefont {C.~H.}\
  \bibnamefont {Keitel}},\ and\ \bibinfo {author} {\bibfnamefont
  {R.}~\bibnamefont {Moshammer}},\ }\bibfield  {title} {\bibinfo {title}
  {Experimental evidence for wigner's tunneling time},\ }\href@noop {}
  {\bibfield  {journal} {\bibinfo  {journal} {Phys. Rev. Lett.}\ }\textbf
  {\bibinfo {volume} {119}},\ \bibinfo {pages} {023201} (\bibinfo {year}
  {2017})}\BibitemShut {NoStop}%
\bibitem [{\citenamefont {Sainadh}\ \emph {et~al.}(2019)\citenamefont
  {Sainadh}, \citenamefont {Xu}, \citenamefont {Wang}, \citenamefont
  {Atia-Tul-Noor}, \citenamefont {Wallace}, \citenamefont {Douguet},
  \citenamefont {Bray}, \citenamefont {Ivanov}, \citenamefont {Bartschat},
  \citenamefont {Kheifets}, \citenamefont {Sang},\ and\ \citenamefont
  {Litvinyuk}}]{Sainadh_2019}%
  \BibitemOpen
  \bibfield  {author} {\bibinfo {author} {\bibfnamefont {U.~S.}\ \bibnamefont
  {Sainadh}}, \bibinfo {author} {\bibfnamefont {H.}~\bibnamefont {Xu}},
  \bibinfo {author} {\bibfnamefont {X.}~\bibnamefont {Wang}}, \bibinfo {author}
  {\bibfnamefont {A.}~\bibnamefont {Atia-Tul-Noor}}, \bibinfo {author}
  {\bibfnamefont {W.~C.}\ \bibnamefont {Wallace}}, \bibinfo {author}
  {\bibfnamefont {N.}~\bibnamefont {Douguet}}, \bibinfo {author} {\bibfnamefont
  {A.}~\bibnamefont {Bray}}, \bibinfo {author} {\bibfnamefont {I.}~\bibnamefont
  {Ivanov}}, \bibinfo {author} {\bibfnamefont {K.}~\bibnamefont {Bartschat}},
  \bibinfo {author} {\bibfnamefont {A.}~\bibnamefont {Kheifets}}, \bibinfo
  {author} {\bibfnamefont {R.~T.}\ \bibnamefont {Sang}},\ and\ \bibinfo
  {author} {\bibfnamefont {I.~V.}\ \bibnamefont {Litvinyuk}},\ }\bibfield
  {title} {\bibinfo {title} {Attosecond angular streaking and tunnelling time
  in atomic hydrogen},\ }\href@noop {} {\bibfield  {journal} {\bibinfo
  {journal} {Nature}\ }\textbf {\bibinfo {volume} {568}},\ \bibinfo {pages}
  {75} (\bibinfo {year} {2019})}\BibitemShut {NoStop}%
\bibitem [{\citenamefont {Ni}\ \emph {et~al.}(2016)\citenamefont {Ni},
  \citenamefont {Saalmann},\ and\ \citenamefont {Rost}}]{Ni_2016}%
  \BibitemOpen
  \bibfield  {author} {\bibinfo {author} {\bibfnamefont {H.}~\bibnamefont
  {Ni}}, \bibinfo {author} {\bibfnamefont {U.}~\bibnamefont {Saalmann}},\ and\
  \bibinfo {author} {\bibfnamefont {J.-M.}\ \bibnamefont {Rost}},\ }\bibfield
  {title} {\bibinfo {title} {Tunneling ionization time resolved by
  backpropagation},\ }\href@noop {} {\bibfield  {journal} {\bibinfo  {journal}
  {Phys. Rev. Lett.}\ }\textbf {\bibinfo {volume} {117}},\ \bibinfo {pages}
  {023002} (\bibinfo {year} {2016})}\BibitemShut {NoStop}%
\bibitem [{\citenamefont {Ni}\ \emph {et~al.}(2018{\natexlab{a}})\citenamefont
  {Ni}, \citenamefont {Saalmann},\ and\ \citenamefont {Rost}}]{Ni_2018b}%
  \BibitemOpen
  \bibfield  {author} {\bibinfo {author} {\bibfnamefont {H.}~\bibnamefont
  {Ni}}, \bibinfo {author} {\bibfnamefont {U.}~\bibnamefont {Saalmann}},\ and\
  \bibinfo {author} {\bibfnamefont {J.-M.}\ \bibnamefont {Rost}},\ }\bibfield
  {title} {\bibinfo {title} {{Tunneling exit characteristics from classical
  backpropagation of an ionized electron wave packet}},\ }\href@noop {}
  {\bibfield  {journal} {\bibinfo  {journal} {Phys. Rev. A}\ }\textbf {\bibinfo
  {volume} {97}},\ \bibinfo {pages} {013426} (\bibinfo {year}
  {2018}{\natexlab{a}})}\BibitemShut {NoStop}%
\bibitem [{\citenamefont {Ni}\ \emph {et~al.}(2018{\natexlab{b}})\citenamefont
  {Ni}, \citenamefont {Eicke}, \citenamefont {Ruiz}, \citenamefont {Cai},
  \citenamefont {Oppermann}, \citenamefont {Shvetsov-Shilovski},\ and\
  \citenamefont {Pi}}]{Ni_2018a}%
  \BibitemOpen
  \bibfield  {author} {\bibinfo {author} {\bibfnamefont {H.}~\bibnamefont
  {Ni}}, \bibinfo {author} {\bibfnamefont {N.}~\bibnamefont {Eicke}}, \bibinfo
  {author} {\bibfnamefont {C.}~\bibnamefont {Ruiz}}, \bibinfo {author}
  {\bibfnamefont {J.}~\bibnamefont {Cai}}, \bibinfo {author} {\bibfnamefont
  {F.}~\bibnamefont {Oppermann}}, \bibinfo {author} {\bibfnamefont {N.~I.}\
  \bibnamefont {Shvetsov-Shilovski}},\ and\ \bibinfo {author} {\bibfnamefont
  {L.-W.}\ \bibnamefont {Pi}},\ }\bibfield  {title} {\bibinfo {title}
  {{Tunneling criteria and a nonadiabatic term for strong-field ionization}},\
  }\href@noop {} {\bibfield  {journal} {\bibinfo  {journal} {Phys. Rev. A}\
  }\textbf {\bibinfo {volume} {98}},\ \bibinfo {pages} {013411} (\bibinfo
  {year} {2018}{\natexlab{b}})}\BibitemShut {NoStop}%
\bibitem [{\citenamefont {Teeny}\ \emph
  {et~al.}(2016{\natexlab{a}})\citenamefont {Teeny}, \citenamefont {Yakaboylu},
  \citenamefont {Bauke},\ and\ \citenamefont {Keitel}}]{Teeny_2016a}%
  \BibitemOpen
  \bibfield  {author} {\bibinfo {author} {\bibfnamefont {N.}~\bibnamefont
  {Teeny}}, \bibinfo {author} {\bibfnamefont {E.}~\bibnamefont {Yakaboylu}},
  \bibinfo {author} {\bibfnamefont {H.}~\bibnamefont {Bauke}},\ and\ \bibinfo
  {author} {\bibfnamefont {C.~H.}\ \bibnamefont {Keitel}},\ }\bibfield  {title}
  {\bibinfo {title} {{Ionization Time and Exit Momentum in Strong-Field Tunnel
  Ionization}},\ }\href@noop {} {\bibfield  {journal} {\bibinfo  {journal}
  {Phys. Rev. Lett.}\ }\textbf {\bibinfo {volume} {116}},\ \bibinfo {pages}
  {063003} (\bibinfo {year} {2016}{\natexlab{a}})}\BibitemShut {NoStop}%
\bibitem [{\citenamefont {Teeny}\ \emph
  {et~al.}(2016{\natexlab{b}})\citenamefont {Teeny}, \citenamefont {Keitel},\
  and\ \citenamefont {Bauke}}]{Teeny_2016b}%
  \BibitemOpen
  \bibfield  {author} {\bibinfo {author} {\bibfnamefont {N.}~\bibnamefont
  {Teeny}}, \bibinfo {author} {\bibfnamefont {C.~H.}\ \bibnamefont {Keitel}},\
  and\ \bibinfo {author} {\bibfnamefont {H.}~\bibnamefont {Bauke}},\ }\bibfield
   {title} {\bibinfo {title} {{Virtual-detector approach to tunnel ionization
  and tunneling times}},\ }\href@noop {} {\bibfield  {journal} {\bibinfo
  {journal} {Phys. Rev. A}\ }\textbf {\bibinfo {volume} {94}},\ \bibinfo
  {pages} {022104} (\bibinfo {year} {2016}{\natexlab{b}})}\BibitemShut
  {NoStop}%
\bibitem [{\citenamefont {Yakaboylu}\ \emph {et~al.}(2013)\citenamefont
  {Yakaboylu}, \citenamefont {Klaiber}, \citenamefont {Bauke}, \citenamefont
  {Hatsagortsyan},\ and\ \citenamefont {Keitel}}]{Yakaboylu_2013}%
  \BibitemOpen
  \bibfield  {author} {\bibinfo {author} {\bibfnamefont {E.}~\bibnamefont
  {Yakaboylu}}, \bibinfo {author} {\bibfnamefont {M.}~\bibnamefont {Klaiber}},
  \bibinfo {author} {\bibfnamefont {H.}~\bibnamefont {Bauke}}, \bibinfo
  {author} {\bibfnamefont {K.~Z.}\ \bibnamefont {Hatsagortsyan}},\ and\
  \bibinfo {author} {\bibfnamefont {C.~H.}\ \bibnamefont {Keitel}},\ }\bibfield
   {title} {\bibinfo {title} {Relativistic features and time delay of
  laser-induced tunnel ionization},\ }\href@noop {} {\bibfield  {journal}
  {\bibinfo  {journal} {Phys. Rev. A}\ }\textbf {\bibinfo {volume} {88}},\
  \bibinfo {pages} {063421} (\bibinfo {year} {2013})}\BibitemShut {NoStop}%
\bibitem [{\citenamefont {Kheifets}(2020)}]{Kheifets_2020}%
  \BibitemOpen
  \bibfield  {author} {\bibinfo {author} {\bibfnamefont {A.~S.}\ \bibnamefont
  {Kheifets}},\ }\bibfield  {title} {\bibinfo {title} {The attoclock and the
  tunneling time debate},\ }\href@noop {} {\bibfield  {journal} {\bibinfo
  {journal} {J. Phys. B}\ }\textbf {\bibinfo {volume} {53}},\ \bibinfo {pages}
  {072001} (\bibinfo {year} {2020})}\BibitemShut {NoStop}%
\bibitem [{\citenamefont {Eicke}\ and\ \citenamefont
  {Lein}(2018)}]{Eicke_2018}%
  \BibitemOpen
  \bibfield  {author} {\bibinfo {author} {\bibfnamefont {N.}~\bibnamefont
  {Eicke}}\ and\ \bibinfo {author} {\bibfnamefont {M.}~\bibnamefont {Lein}},\
  }\bibfield  {title} {\bibinfo {title} {{Trajectory-free ionization times in
  strong-field ionization}},\ }\href@noop {} {\bibfield  {journal} {\bibinfo
  {journal} {Phys. Rev. A}\ }\textbf {\bibinfo {volume} {97}},\ \bibinfo
  {pages} {031402} (\bibinfo {year} {2018})}\BibitemShut {NoStop}%
\bibitem [{\citenamefont {Douguet}\ and\ \citenamefont
  {Bartschat}(2018)}]{Douguet_2018}%
  \BibitemOpen
  \bibfield  {author} {\bibinfo {author} {\bibfnamefont {N.}~\bibnamefont
  {Douguet}}\ and\ \bibinfo {author} {\bibfnamefont {K.}~\bibnamefont
  {Bartschat}},\ }\bibfield  {title} {\bibinfo {title} {{Dynamics of Tunneling
  Ionization using Bohmian Mechanics}},\ }\href@noop {} {\bibfield  {journal}
  {\bibinfo  {journal} {Phys. Rev. A}\ }\textbf {\bibinfo {volume} {97}},\
  \bibinfo {pages} {013402} (\bibinfo {year} {2018})}\BibitemShut {NoStop}%
\bibitem [{\citenamefont {Bray}\ \emph {et~al.}(2018)\citenamefont {Bray},
  \citenamefont {Eckart},\ and\ \citenamefont {Kheifets}}]{Bray_2018}%
  \BibitemOpen
  \bibfield  {author} {\bibinfo {author} {\bibfnamefont {A.~W.}\ \bibnamefont
  {Bray}}, \bibinfo {author} {\bibfnamefont {S.}~\bibnamefont {Eckart}},\ and\
  \bibinfo {author} {\bibfnamefont {A.~S.}\ \bibnamefont {Kheifets}},\
  }\bibfield  {title} {\bibinfo {title} {Keldysh-rutherford model for the
  attoclock},\ }\href@noop {} {\bibfield  {journal} {\bibinfo  {journal} {Phys.
  Rev. Lett.}\ }\textbf {\bibinfo {volume} {121}},\ \bibinfo {pages} {123201}
  (\bibinfo {year} {2018})}\BibitemShut {NoStop}%
\bibitem [{\citenamefont {Can\'ario}\ \emph {et~al.}(2021)\citenamefont
  {Can\'ario}, \citenamefont {Klaiber}, \citenamefont {Hatsagortsyan},\ and\
  \citenamefont {Keitel}}]{Canario_2021}%
  \BibitemOpen
  \bibfield  {author} {\bibinfo {author} {\bibfnamefont {D.~B.}\ \bibnamefont
  {Can\'ario}}, \bibinfo {author} {\bibfnamefont {M.}~\bibnamefont {Klaiber}},
  \bibinfo {author} {\bibfnamefont {K.~Z.}\ \bibnamefont {Hatsagortsyan}},\
  and\ \bibinfo {author} {\bibfnamefont {C.~H.}\ \bibnamefont {Keitel}},\
  }\bibfield  {title} {\bibinfo {title} {Role of reflections in the generation
  of a time delay in strong-field ionization},\ }\href
  {https://doi.org/10.1103/PhysRevA.104.033103} {\bibfield  {journal} {\bibinfo
   {journal} {Phys. Rev. A}\ }\textbf {\bibinfo {volume} {104}},\ \bibinfo
  {pages} {033103} (\bibinfo {year} {2021})}\BibitemShut {NoStop}%
\bibitem [{\citenamefont {Klaiber}\ \emph
  {et~al.}(2022{\natexlab{a}})\citenamefont {Klaiber}, \citenamefont {Lv},
  \citenamefont {Sukiasyan}, \citenamefont {Bakucz~Can\'ario}, \citenamefont
  {Hatsagortsyan},\ and\ \citenamefont {Keitel}}]{Klaiber_2022R}%
  \BibitemOpen
  \bibfield  {author} {\bibinfo {author} {\bibfnamefont {M.}~\bibnamefont
  {Klaiber}}, \bibinfo {author} {\bibfnamefont {Q.~Z.}\ \bibnamefont {Lv}},
  \bibinfo {author} {\bibfnamefont {S.}~\bibnamefont {Sukiasyan}}, \bibinfo
  {author} {\bibfnamefont {D.}~\bibnamefont {Bakucz~Can\'ario}}, \bibinfo
  {author} {\bibfnamefont {K.~Z.}\ \bibnamefont {Hatsagortsyan}},\ and\
  \bibinfo {author} {\bibfnamefont {C.~H.}\ \bibnamefont {Keitel}},\ }\bibfield
   {title} {\bibinfo {title} {Reconciling conflicting approaches for the
  tunneling time delay in strong field ionization},\ }\href
  {https://doi.org/10.1103/PhysRevLett.129.203201} {\bibfield  {journal}
  {\bibinfo  {journal} {Phys. Rev. Lett.}\ }\textbf {\bibinfo {volume} {129}},\
  \bibinfo {pages} {203201} (\bibinfo {year} {2022}{\natexlab{a}})}\BibitemShut
  {NoStop}%
\bibitem [{\citenamefont {Klaiber}\ \emph {et~al.}(2023)\citenamefont
  {Klaiber}, \citenamefont {Bakucz~Can\'ario},\ and\ \citenamefont
  {Hatsagortsyan}}]{Klaiber_2023}%
  \BibitemOpen
  \bibfield  {author} {\bibinfo {author} {\bibfnamefont {M.}~\bibnamefont
  {Klaiber}}, \bibinfo {author} {\bibfnamefont {D.}~\bibnamefont
  {Bakucz~Can\'ario}},\ and\ \bibinfo {author} {\bibfnamefont {K.~Z.}\
  \bibnamefont {Hatsagortsyan}},\ }\bibfield  {title} {\bibinfo {title}
  {Sub-barrier recollisions and the three classes of tunneling time delays in
  strong-field ionization},\ }\href
  {https://doi.org/10.1103/PhysRevA.107.053103} {\bibfield  {journal} {\bibinfo
   {journal} {Phys. Rev. A}\ }\textbf {\bibinfo {volume} {107}},\ \bibinfo
  {pages} {053103} (\bibinfo {year} {2023})}\BibitemShut {NoStop}%
\bibitem [{\citenamefont {Klaiber}\ \emph {et~al.}(2024)\citenamefont
  {Klaiber}, \citenamefont {Hatsagortsyan},\ and\ \citenamefont
  {Keitel}}]{Klaiber_2024}%
  \BibitemOpen
  \bibfield  {author} {\bibinfo {author} {\bibfnamefont {M.}~\bibnamefont
  {Klaiber}}, \bibinfo {author} {\bibfnamefont {K.~Z.}\ \bibnamefont
  {Hatsagortsyan}},\ and\ \bibinfo {author} {\bibfnamefont {C.~H.}\
  \bibnamefont {Keitel}},\ }\bibfield  {title} {\bibinfo {title} {Signatures of
  under-the-barrier dynamics in a tunneling electron wavepacket},\ }\href@noop
  {} {\bibfield  {journal} {\bibinfo  {journal} {Commun Phys}\ }\textbf
  {\bibinfo {volume} {7}},\ \bibinfo {pages} {384} (\bibinfo {year}
  {2024})}\BibitemShut {NoStop}%
\bibitem [{\citenamefont {Freeman}\ \emph {et~al.}(1987)\citenamefont
  {Freeman}, \citenamefont {Bucksbaum}, \citenamefont {Milchberg},
  \citenamefont {Darack}, \citenamefont {Schumacher},\ and\ \citenamefont
  {Geusic}}]{Freeman_1987}%
  \BibitemOpen
  \bibfield  {author} {\bibinfo {author} {\bibfnamefont {R.~R.}\ \bibnamefont
  {Freeman}}, \bibinfo {author} {\bibfnamefont {P.~H.}\ \bibnamefont
  {Bucksbaum}}, \bibinfo {author} {\bibfnamefont {H.}~\bibnamefont
  {Milchberg}}, \bibinfo {author} {\bibfnamefont {S.}~\bibnamefont {Darack}},
  \bibinfo {author} {\bibfnamefont {D.}~\bibnamefont {Schumacher}},\ and\
  \bibinfo {author} {\bibfnamefont {M.~E.}\ \bibnamefont {Geusic}},\ }\bibfield
   {title} {\bibinfo {title} {{Above-threshold ionization with subpicosecond
  laser pulses}},\ }\href@noop {} {\bibfield  {journal} {\bibinfo  {journal}
  {Phys. Rev. Lett.}\ }\textbf {\bibinfo {volume} {59}},\ \bibinfo {pages}
  {1092} (\bibinfo {year} {1987})}\BibitemShut {NoStop}%
\bibitem [{\citenamefont {Klaiber}\ \emph {et~al.}(2020)\citenamefont
  {Klaiber}, \citenamefont {Hatsagortsyan},\ and\ \citenamefont
  {Keitel}}]{Klaiber_2020}%
  \BibitemOpen
  \bibfield  {author} {\bibinfo {author} {\bibfnamefont {M.}~\bibnamefont
  {Klaiber}}, \bibinfo {author} {\bibfnamefont {K.~Z.}\ \bibnamefont
  {Hatsagortsyan}},\ and\ \bibinfo {author} {\bibfnamefont {C.~H.}\
  \bibnamefont {Keitel}},\ }\bibfield  {title} {\bibinfo {title} {{Sub-barrier
  pathways to Freeman resonances}},\ }\href
  {https://doi.org/10.1103/PhysRevA.102.053105} {\bibfield  {journal} {\bibinfo
   {journal} {Phys. Rev. A}\ }\textbf {\bibinfo {volume} {102}},\ \bibinfo
  {pages} {053105} (\bibinfo {year} {2020})}\BibitemShut {NoStop}%
\bibitem [{\citenamefont {Khurelbaatar}\ \emph {et~al.}(2025)\citenamefont
  {Khurelbaatar}, \citenamefont {Klaiber}, \citenamefont {Sukiasyan},
  \citenamefont {Hatsagortsyan}, \citenamefont {Keitel},\ and\ \citenamefont
  {Kim}}]{Khurelbaatar_2025}%
  \BibitemOpen
  \bibfield  {author} {\bibinfo {author} {\bibfnamefont {T.}~\bibnamefont
  {Khurelbaatar}}, \bibinfo {author} {\bibfnamefont {M.}~\bibnamefont
  {Klaiber}}, \bibinfo {author} {\bibfnamefont {S.}~\bibnamefont {Sukiasyan}},
  \bibinfo {author} {\bibfnamefont {K.~Z.}\ \bibnamefont {Hatsagortsyan}},
  \bibinfo {author} {\bibfnamefont {C.~H.}\ \bibnamefont {Keitel}},\ and\
  \bibinfo {author} {\bibfnamefont {D.~E.}\ \bibnamefont {Kim}},\ }\bibfield
  {title} {\bibinfo {title} {Unveiling under-the-barrier electron dynamics in
  strong field tunneling},\ }\href
  {https://doi.org/10.1103/PhysRevLett.134.213201} {\bibfield  {journal}
  {\bibinfo  {journal} {Phys. Rev. Lett.}\ }\textbf {\bibinfo {volume} {134}},\
  \bibinfo {pages} {213201} (\bibinfo {year} {2025})}\BibitemShut {NoStop}%
\bibitem [{\citenamefont {Faisal}(1973)}]{Faisal_1973}%
  \BibitemOpen
  \bibfield  {author} {\bibinfo {author} {\bibfnamefont {F.~H.~M.}\
  \bibnamefont {Faisal}},\ }\bibfield  {title} {\bibinfo {title} {Multiple
  absorption of laser photons by atoms},\ }\href@noop {} {\bibfield  {journal}
  {\bibinfo  {journal} {J. Phys. B}\ }\textbf {\bibinfo {volume} {6}},\
  \bibinfo {pages} {L89} (\bibinfo {year} {1973})}\BibitemShut {NoStop}%
\bibitem [{\citenamefont {Reiss}(1980)}]{Reiss_1980}%
  \BibitemOpen
  \bibfield  {author} {\bibinfo {author} {\bibfnamefont {H.~R.}\ \bibnamefont
  {Reiss}},\ }\bibfield  {title} {\bibinfo {title} {Effect of an intesne
  electromagnetic field on a weakly bound system},\ }\href@noop {} {\bibfield
  {journal} {\bibinfo  {journal} {Phys. Rev. A}\ }\textbf {\bibinfo {volume}
  {22}},\ \bibinfo {pages} {1786} (\bibinfo {year} {1980})}\BibitemShut
  {NoStop}%
\bibitem [{\citenamefont {Yudin}\ and\ \citenamefont
  {Ivanov}(2001)}]{Yudin_2001b}%
  \BibitemOpen
  \bibfield  {author} {\bibinfo {author} {\bibfnamefont {G.~L.}\ \bibnamefont
  {Yudin}}\ and\ \bibinfo {author} {\bibfnamefont {M.~Y.}\ \bibnamefont
  {Ivanov}},\ }\bibfield  {title} {\bibinfo {title} {Nonadiabatic tunnel
  ionization: Looking inside a laser cycle},\ }\href@noop {} {\bibfield
  {journal} {\bibinfo  {journal} {Phys. Rev. A}\ }\textbf {\bibinfo {volume}
  {64}},\ \bibinfo {pages} {013409} (\bibinfo {year} {2001})}\BibitemShut
  {NoStop}%
\bibitem [{\citenamefont {Kaushal}\ and\ \citenamefont
  {Smirnova}(2013)}]{Kaushal_2013}%
  \BibitemOpen
  \bibfield  {author} {\bibinfo {author} {\bibfnamefont {J.}~\bibnamefont
  {Kaushal}}\ and\ \bibinfo {author} {\bibfnamefont {O.}~\bibnamefont
  {Smirnova}},\ }\bibfield  {title} {\bibinfo {title} {Nonadiabatic coulomb
  effects in strong-field ionization in circularly polarized laser fields},\
  }\href@noop {} {\bibfield  {journal} {\bibinfo  {journal} {Phys. Rev. A}\
  }\textbf {\bibinfo {volume} {88}},\ \bibinfo {pages} {013421} (\bibinfo
  {year} {2013})}\BibitemShut {NoStop}%
\bibitem [{\citenamefont {Boge}\ \emph {et~al.}(2013)\citenamefont {Boge},
  \citenamefont {Cirelli}, \citenamefont {Landsman}, \citenamefont {Heuser},
  \citenamefont {Ludwig}, \citenamefont {Maurer}, \citenamefont {Weger},
  \citenamefont {Gallmann},\ and\ \citenamefont {Keller}}]{Boge_2013}%
  \BibitemOpen
  \bibfield  {author} {\bibinfo {author} {\bibfnamefont {R.}~\bibnamefont
  {Boge}}, \bibinfo {author} {\bibfnamefont {C.}~\bibnamefont {Cirelli}},
  \bibinfo {author} {\bibfnamefont {A.~S.}\ \bibnamefont {Landsman}}, \bibinfo
  {author} {\bibfnamefont {S.}~\bibnamefont {Heuser}}, \bibinfo {author}
  {\bibfnamefont {A.}~\bibnamefont {Ludwig}}, \bibinfo {author} {\bibfnamefont
  {J.}~\bibnamefont {Maurer}}, \bibinfo {author} {\bibfnamefont
  {M.}~\bibnamefont {Weger}}, \bibinfo {author} {\bibfnamefont
  {L.}~\bibnamefont {Gallmann}},\ and\ \bibinfo {author} {\bibfnamefont
  {U.}~\bibnamefont {Keller}},\ }\bibfield  {title} {\bibinfo {title} {Probing
  nonadiabatic effects in strong-field tunnel ionization},\ }\href@noop {}
  {\bibfield  {journal} {\bibinfo  {journal} {Phys. Rev. Lett.}\ }\textbf
  {\bibinfo {volume} {111}},\ \bibinfo {pages} {103003} (\bibinfo {year}
  {2013})}\BibitemShut {NoStop}%
\bibitem [{\citenamefont {Klaiber}\ \emph {et~al.}(2015)\citenamefont
  {Klaiber}, , \citenamefont {Hatsagortsyan},\ and\ \citenamefont
  {Keitel}}]{Klaiber_2015}%
  \BibitemOpen
  \bibfield  {author} {\bibinfo {author} {\bibfnamefont {M.}~\bibnamefont
  {Klaiber}}, , \bibinfo {author} {\bibfnamefont {K.~Z.}\ \bibnamefont
  {Hatsagortsyan}},\ and\ \bibinfo {author} {\bibfnamefont {C.~H.}\
  \bibnamefont {Keitel}},\ }\bibfield  {title} {\bibinfo {title} {Tunneling
  dynamics in multiphoton ionization and attoclock calibration},\ }\href@noop
  {} {\bibfield  {journal} {\bibinfo  {journal} {Phys. Rev. Lett.}\ }\textbf
  {\bibinfo {volume} {114}},\ \bibinfo {pages} {083001} (\bibinfo {year}
  {2015})}\BibitemShut {NoStop}%
\bibitem [{\citenamefont {Dubois}\ \emph {et~al.}(2020)\citenamefont {Dubois},
  \citenamefont {Chandre},\ and\ \citenamefont {Uzer}}]{Dubois_2020a}%
  \BibitemOpen
  \bibfield  {author} {\bibinfo {author} {\bibfnamefont {J.}~\bibnamefont
  {Dubois}}, \bibinfo {author} {\bibfnamefont {C.}~\bibnamefont {Chandre}},\
  and\ \bibinfo {author} {\bibfnamefont {T.}~\bibnamefont {Uzer}},\ }\bibfield
  {title} {\bibinfo {title} {Nonadiabatic effects in the double ionization of
  atoms driven by a circularly polarized laser pulse},\ }\href
  {https://doi.org/10.1103/PhysRevE.102.032218} {\bibfield  {journal} {\bibinfo
   {journal} {Phys. Rev. E}\ }\textbf {\bibinfo {volume} {102}},\ \bibinfo
  {pages} {032218} (\bibinfo {year} {2020})}\BibitemShut {NoStop}%
\bibitem [{\citenamefont {Trabert}\ \emph {et~al.}(2021)\citenamefont
  {Trabert}, \citenamefont {Anders}, \citenamefont {Brennecke}, \citenamefont
  {Sch\"offler}, \citenamefont {Jahnke}, \citenamefont {Schmidt}, \citenamefont
  {Kunitski}, \citenamefont {Lein}, \citenamefont {D\"orner},\ and\
  \citenamefont {Eckart}}]{Trabert_2021}%
  \BibitemOpen
  \bibfield  {author} {\bibinfo {author} {\bibfnamefont {D.}~\bibnamefont
  {Trabert}}, \bibinfo {author} {\bibfnamefont {N.}~\bibnamefont {Anders}},
  \bibinfo {author} {\bibfnamefont {S.}~\bibnamefont {Brennecke}}, \bibinfo
  {author} {\bibfnamefont {M.~S.}\ \bibnamefont {Sch\"offler}}, \bibinfo
  {author} {\bibfnamefont {T.}~\bibnamefont {Jahnke}}, \bibinfo {author}
  {\bibfnamefont {L.~P.~H.}\ \bibnamefont {Schmidt}}, \bibinfo {author}
  {\bibfnamefont {M.}~\bibnamefont {Kunitski}}, \bibinfo {author}
  {\bibfnamefont {M.}~\bibnamefont {Lein}}, \bibinfo {author} {\bibfnamefont
  {R.}~\bibnamefont {D\"orner}},\ and\ \bibinfo {author} {\bibfnamefont
  {S.}~\bibnamefont {Eckart}},\ }\bibfield  {title} {\bibinfo {title}
  {Nonadiabatic strong field ionization of atomic hydrogen},\ }\href
  {https://doi.org/10.1103/PhysRevLett.127.273201} {\bibfield  {journal}
  {\bibinfo  {journal} {Phys. Rev. Lett.}\ }\textbf {\bibinfo {volume} {127}},\
  \bibinfo {pages} {273201} (\bibinfo {year} {2021})}\BibitemShut {NoStop}%
\bibitem [{\citenamefont {Walker}\ \emph {et~al.}(2026)\citenamefont {Walker},
  \citenamefont {Alqasem}, \citenamefont {Garibay}, \citenamefont {Blaga},
  \citenamefont {Landsman},\ and\ \citenamefont {DiMauro}}]{Spencer_2026}%
  \BibitemOpen
  \bibfield  {author} {\bibinfo {author} {\bibfnamefont {S.}~\bibnamefont
  {Walker}}, \bibinfo {author} {\bibfnamefont {A.}~\bibnamefont {Alqasem}},
  \bibinfo {author} {\bibfnamefont {A.~C.}\ \bibnamefont {Garibay}}, \bibinfo
  {author} {\bibfnamefont {C.~I.}\ \bibnamefont {Blaga}}, \bibinfo {author}
  {\bibfnamefont {A.~S.}\ \bibnamefont {Landsman}},\ and\ \bibinfo {author}
  {\bibfnamefont {L.~F.}\ \bibnamefont {DiMauro}},\ }\bibfield  {title}
  {\bibinfo {title} {Nonadiabatic strong-field photoionization revisited},\
  }\href@noop {} {\bibfield  {journal} {\bibinfo  {journal} {Phys. Rev. Lett.}\
  }\textbf {\bibinfo {volume} {136}},\ \bibinfo {pages} {123203} (\bibinfo
  {year} {2026})}\BibitemShut {NoStop}%
\bibitem [{\citenamefont {Popruzhenko}(2014)}]{Popruzhenko_2014r}%
  \BibitemOpen
  \bibfield  {author} {\bibinfo {author} {\bibfnamefont {S.~V.}\ \bibnamefont
  {Popruzhenko}},\ }\bibfield  {title} {\bibinfo {title} {{Keldysh theory of
  strong field ionization: history, applications, difficulties and
  perspectives}},\ }\href@noop {} {\bibfield  {journal} {\bibinfo  {journal}
  {J. Phys. B}\ }\textbf {\bibinfo {volume} {47}},\ \bibinfo {pages} {204001}
  (\bibinfo {year} {2014})}\BibitemShut {NoStop}%
\bibitem [{\citenamefont {Becker}\ \emph {et~al.}(2002)\citenamefont {Becker},
  \citenamefont {Grasbon}, \citenamefont {Kopold}, \citenamefont
  {Milo\u{s}evi\'c}, \citenamefont {Paulus},\ and\ \citenamefont
  {Walther}}]{Becker_2002}%
  \BibitemOpen
  \bibfield  {author} {\bibinfo {author} {\bibfnamefont {W.}~\bibnamefont
  {Becker}}, \bibinfo {author} {\bibfnamefont {F.}~\bibnamefont {Grasbon}},
  \bibinfo {author} {\bibfnamefont {R.}~\bibnamefont {Kopold}}, \bibinfo
  {author} {\bibfnamefont {D.~B.}\ \bibnamefont {Milo\u{s}evi\'c}}, \bibinfo
  {author} {\bibfnamefont {G.~G.}\ \bibnamefont {Paulus}},\ and\ \bibinfo
  {author} {\bibfnamefont {H.}~\bibnamefont {Walther}},\ }\bibfield  {title}
  {\bibinfo {title} {Above-threshold ionization: from classical features to
  quantum effects},\ }\href@noop {} {\bibfield  {journal} {\bibinfo  {journal}
  {Adv. Atom. Mol. Opt. Phys.}\ }\textbf {\bibinfo {volume} {48}},\ \bibinfo
  {pages} {35} (\bibinfo {year} {2002})}\BibitemShut {NoStop}%
\bibitem [{\citenamefont {Klaiber}\ \emph
  {et~al.}(2022{\natexlab{b}})\citenamefont {Klaiber}, \citenamefont {Lv},
  \citenamefont {Hatsagortsyan},\ and\ \citenamefont
  {Keitel}}]{Klaiber_2022edge}%
  \BibitemOpen
  \bibfield  {author} {\bibinfo {author} {\bibfnamefont {M.}~\bibnamefont
  {Klaiber}}, \bibinfo {author} {\bibfnamefont {Q.~Z.}\ \bibnamefont {Lv}},
  \bibinfo {author} {\bibfnamefont {K.~Z.}\ \bibnamefont {Hatsagortsyan}},\
  and\ \bibinfo {author} {\bibfnamefont {C.~H.}\ \bibnamefont {Keitel}},\
  }\bibfield  {title} {\bibinfo {title} {Tunneling ionization in ultrashort
  laser pulses: Edge effect and remedy},\ }\href
  {https://doi.org/10.1103/PhysRevA.105.063109} {\bibfield  {journal} {\bibinfo
   {journal} {Phys. Rev. A}\ }\textbf {\bibinfo {volume} {105}},\ \bibinfo
  {pages} {063109} (\bibinfo {year} {2022}{\natexlab{b}})}\BibitemShut
  {NoStop}%
\bibitem [{\citenamefont {Klaiber}\ \emph {et~al.}(2026)\citenamefont
  {Klaiber}, \citenamefont {Hatsagortsyan},\ and\ \citenamefont
  {Keitel}}]{Klaiber_2026}%
  \BibitemOpen
  \bibfield  {author} {\bibinfo {author} {\bibfnamefont {M.}~\bibnamefont
  {Klaiber}}, \bibinfo {author} {\bibfnamefont {K.~Z.}\ \bibnamefont
  {Hatsagortsyan}},\ and\ \bibinfo {author} {\bibfnamefont {C.~H.}\
  \bibnamefont {Keitel}},\ }\bibfield  {title} {\bibinfo {title} {Role of
  quantum dynamics in coherent and incoherent radiation during tunneling
  ionization},\ }\href@noop {} {\bibfield  {journal} {\bibinfo  {journal}
  {Phys. Rev. A}\ }\textbf {\bibinfo {volume} {113}},\ \bibinfo {pages}
  {033110} (\bibinfo {year} {2026})}\BibitemShut {NoStop}%
\bibitem [{\citenamefont {\ifmmode \check{C}\else
  \v{C}\fi{}erki\ifmmode~\acute{c}\else \'{c}\fi{}}\ \emph
  {et~al.}(2009)\citenamefont {\ifmmode \check{C}\else
  \v{C}\fi{}erki\ifmmode~\acute{c}\else \'{c}\fi{}}, \citenamefont
  {Hasovi\ifmmode~\acute{c}\else \'{c}\fi{}}, \citenamefont {Milo\ifmmode
  \check{s}\else \v{s}\fi{}evi\ifmmode~\acute{c}\else \'{c}\fi{}},\ and\
  \citenamefont {Becker}}]{Cerkic_2009}%
  \BibitemOpen
  \bibfield  {author} {\bibinfo {author} {\bibfnamefont {A.}~\bibnamefont
  {\ifmmode \check{C}\else \v{C}\fi{}erki\ifmmode~\acute{c}\else \'{c}\fi{}}},
  \bibinfo {author} {\bibfnamefont {E.}~\bibnamefont
  {Hasovi\ifmmode~\acute{c}\else \'{c}\fi{}}}, \bibinfo {author} {\bibfnamefont
  {D.~B.}\ \bibnamefont {Milo\ifmmode \check{s}\else
  \v{s}\fi{}evi\ifmmode~\acute{c}\else \'{c}\fi{}}},\ and\ \bibinfo {author}
  {\bibfnamefont {W.}~\bibnamefont {Becker}},\ }\bibfield  {title} {\bibinfo
  {title} {High-order above-threshold ionization beyond the first-order born
  approximation},\ }\href@noop {} {\bibfield  {journal} {\bibinfo  {journal}
  {Phys. Rev. A}\ }\textbf {\bibinfo {volume} {79}},\ \bibinfo {pages} {033413}
  (\bibinfo {year} {2009})}\BibitemShut {NoStop}%
\bibitem [{\citenamefont {Milo\ifmmode \check{s}\else
  \v{s}\fi{}evi\ifmmode~\acute{c}\else \'{c}\fi{}}(2014)}]{Milosevic_2014}%
  \BibitemOpen
  \bibfield  {author} {\bibinfo {author} {\bibfnamefont {D.~B.}\ \bibnamefont
  {Milo\ifmmode \check{s}\else \v{s}\fi{}evi\ifmmode~\acute{c}\else
  \'{c}\fi{}}},\ }\bibfield  {title} {\bibinfo {title} {Forward- and
  backward-scattering quantum orbits in above-threshold ionization},\
  }\href@noop {} {\bibfield  {journal} {\bibinfo  {journal} {Phys. Rev. A}\
  }\textbf {\bibinfo {volume} {90}},\ \bibinfo {pages} {063414} (\bibinfo
  {year} {2014})}\BibitemShut {NoStop}%
\bibitem [{\citenamefont {Lohr}\ \emph {et~al.}(1997)\citenamefont {Lohr},
  \citenamefont {Kleber}, \citenamefont {Kopold},\ and\ \citenamefont
  {Becker}}]{Lohr_1997}%
  \BibitemOpen
  \bibfield  {author} {\bibinfo {author} {\bibfnamefont {A.}~\bibnamefont
  {Lohr}}, \bibinfo {author} {\bibfnamefont {M.}~\bibnamefont {Kleber}},
  \bibinfo {author} {\bibfnamefont {R.}~\bibnamefont {Kopold}},\ and\ \bibinfo
  {author} {\bibfnamefont {W.}~\bibnamefont {Becker}},\ }\bibfield  {title}
  {\bibinfo {title} {Above-threshold ionization in the tunneling regime},\
  }\href@noop {} {\bibfield  {journal} {\bibinfo  {journal} {Phys. Rev. A}\
  }\textbf {\bibinfo {volume} {55}},\ \bibinfo {pages} {R4003} (\bibinfo {year}
  {1997})}\BibitemShut {NoStop}%
\bibitem [{\citenamefont {Krajewska}\ \emph {et~al.}(2010)\citenamefont
  {Krajewska}, \citenamefont {Kami\'nski},\ and\ \citenamefont
  {W\'odkiewicz}}]{Krajewska_2010}%
  \BibitemOpen
  \bibfield  {author} {\bibinfo {author} {\bibfnamefont {K.}~\bibnamefont
  {Krajewska}}, \bibinfo {author} {\bibfnamefont {J.}~\bibnamefont
  {Kami\'nski}},\ and\ \bibinfo {author} {\bibfnamefont {K.}~\bibnamefont
  {W\'odkiewicz}},\ }\bibfield  {title} {\bibinfo {title} {Zero-range
  interaction in arbitrary dimensions and in the presence of external forces},\
  }\href@noop {} {\bibfield  {journal} {\bibinfo  {journal} {Opt. Commun.}\
  }\textbf {\bibinfo {volume} {283}},\ \bibinfo {pages} {843 } (\bibinfo {year}
  {2010})}\BibitemShut {NoStop}%
\bibitem [{\citenamefont {Tolstikhin}\ and\ \citenamefont
  {Madsen}(2013)}]{Tolstikhin_2013}%
  \BibitemOpen
  \bibfield  {author} {\bibinfo {author} {\bibfnamefont {O.~I.}\ \bibnamefont
  {Tolstikhin}}\ and\ \bibinfo {author} {\bibfnamefont {L.~B.}\ \bibnamefont
  {Madsen}},\ }\bibfield  {title} {\bibinfo {title} {Retardation effects and
  the born-oppenheimer approximation: Theory of tunneling ionization of
  molecules revisited},\ }\href@noop {} {\bibfield  {journal} {\bibinfo
  {journal} {Phys. Rev. Lett.}\ }\textbf {\bibinfo {volume} {111}},\ \bibinfo
  {pages} {153003} (\bibinfo {year} {2013})}\BibitemShut {NoStop}%
\end{thebibliography}%

\end{document}